\documentclass[aps,prd,twocolumn,superscriptaddress,nolongbibliography,nofootinbib]{revtex4-2}

\usepackage{xcolor}
\usepackage{graphicx}
\usepackage{amsmath}

\colorlet{darkBlue}{blue!50!black}
\usepackage[colorlinks=true, pdfborder={0 0 0}, linkcolor=darkBlue, citecolor=darkBlue, urlcolor=black, bookmarks, pdftitle={}]{hyperref}

\begin{document}

\title{Constraining the contribution of Seyfert galaxies to the diffuse neutrino flux in light of point source observations}

% author list
\author{Lena Saurenhaus}
\email[]{lsaurenh@mpp.mpg.de}
\affiliation{Max Planck Institute for Physics, 85748 Garching, Germany}
\affiliation{Department of Physics, TUM School of Natural Sciences, Technical University of Munich, 85747 Garching, Germany}

\author{Francesca Capel}
\affiliation{Max Planck Institute for Physics, 85748 Garching, Germany}

\author{Foteini Oikonomou}
\affiliation{Institutt for fysikk, Norwegian University of Science and Technology, 7491 Trondheim, Norway}

\author{Johannes Buchner}
\affiliation{Max Planck Institute for Extraterrestrial Physics, 85748 Garching, Germany}

\date{\today}

% -------------------------------------------------------------------------------------------------

\begin{abstract}

Recently, the IceCube Collaboration reported evidence for TeV neutrino emission from several nearby Seyfert galaxies, with the highest significance found for NGC 1068. Assuming stochastic proton acceleration in magnetized turbulence inside the corona, we model the neutrino emission of Seyfert galaxies as a function of their X-ray luminosity. Applying our model to NGC 1068, we obtain a good fit to the public IceCube data and constrain the coronal radius to $\lesssim 5 R_S$ by comparing our MeV $\gamma$-ray predictions to Fermi-LAT observations. Extending to the full Seyfert population, we estimate their diffuse neutrino contribution and find that they can explain a significant fraction of the observed flux below $10\,\mathrm{TeV}$. However, scenarios with highly turbulent coronae and high cosmic-ray pressure across the population are ruled out. In particular, if all sources shared the best-fit parameters obtained for NGC 1068, their cumulative neutrino emission would exceed current upper limits at TeV energies by $3.8\sigma$. Our results, informed by both neutrino and $\gamma$-ray data, show that those Seyfert galaxies that emerge as neutrino point sources must be exceptionally efficient neutrino emitters and are not representative of the broader population.

\end{abstract}

\maketitle

% -------------------------------------------------------------------------------------------------

\section{Introduction}

The discovery of a diffuse flux of astrophysical neutrinos by the IceCube Neutrino Observatory has provided unambiguous evidence for the existence of powerful cosmic accelerators capable of producing neutrinos up to $\sim\,$PeV energies \cite{Aartsen2013a, Aartsen2013b}. Yet, more than a decade later, identifying the origins of these neutrinos remains challenging. 

The recent IceCube observation of an excess of neutrinos from the direction of the nearby Seyfert galaxy NGC~1068 is a promising development \cite{Abbasi2022}. Evidence is building for neutrino emission from this source class, with further studies by the IceCube Collaboration showing possible signals from other Seyfert galaxies, particularly NGC~4151 and CGCG~420-015 \cite{Abbasi2024a, Abbasi2025}. Independently, a search using public IceCube data identified NGC~4151 and NGC~3079 as possible neutrino sources \cite{Neronov2024}, while \cite{Sommani2025} highlighted the coincidence of two $\sim\,$100$\, \mathrm{TeV}$ IceCube alert events with the location of the Seyfert galaxy NGC~7469. 

NGC 1068 remains the strongest source in the Northern sky, with a significance of $4.2\sigma$. As such, its potential as a neutrino emitter has been extensively studied from both observational and theoretical perspectives (see Ref.~\cite{Padovani2024a} and references therein). Located at a distance of $d_L = 11.14\, \mathrm{Mpc}$ \cite{Tikhonov2021}, NGC~1068 hosts a luminous obscured active nucleus with an intrinsic X-ray luminosity of $L_X = 4.17 \times 10^{43}\, \mathrm{erg}\, \mathrm{s}^{-1}$ in the 2–10 keV range (after rescaling the measurement of~\cite{Marinucci2016} to the updated luminosity distance). This makes it the Seyfert galaxy with the highest intrinsic 2–10 keV X-ray flux in the Northern Hemisphere \cite{Ricci2017}.

Notably, no TeV $\gamma$-ray emission has been detected from NGC~1068 despite searches by ground-based instruments such as HESS \cite{Aharonian2005}, MAGIC \cite{Acciari2019}, or HAWC \cite{Albert2021}. This non-detection implies that any $\gamma$-rays produced in the same hadronic processes responsible for neutrino emission must be efficiently absorbed within the source. NGC~1068 thus represents a compelling example of a ``hidden'' neutrino source, a class that has been independently proposed to explain the tension between measurements of the diffuse $\gamma$-ray and $< 10\, \mathrm{TeV}$ neutrino fluxes~\cite{Murase:2015xka}. 

Several theoretical models suggest that the observed neutrinos associated with NGC 1068 originate from hadronic interactions in a dense corona located close to the central supermassive black hole (SMBH). Murase \textit{et al.}~\cite{Murase2020} propose that inside this corona, protons are accelerated via stochastic acceleration in magnetized turbulence and subsequently interact with ambient gas and coronal X-rays, producing high-energy neutrinos and $\gamma$-rays. The latter are attenuated in $\gamma\gamma$ interactions with X-rays, making it possible to explain the observed neutrino emission without violating any $\gamma$-ray observations. Building a two-zone model of the corona and the circumnuclear starburst region of NGC 1068, Eichmann \textit{et al.}~\cite{Eichmann2022} confirm that the neutrinos can only be produced inside the corona, while the starburst region contributes to the $\gamma$-ray emission. Based on radio observations of an excess in the mm-range, which can be interpreted as coronal synchrotron emission, Inoue \textit{et al.}~\cite{Inoue2019, Inoue2020} assume that both electrons and protons are accelerated via diffusive shock acceleration. In addition, magnetic reconnection inside the corona has been proposed as a possible proton acceleration scenario \cite{Kheirandish2021, Fiorillo2024, Karavola2025}.

Alternative scenarios involve outflows driven by the active galactic nucleus (AGN) \cite{Lamastra2016, Peretti2023, Inoue2022}, jet interactions with the interstellar medium \cite{Fang2023}, or stellar-mass black holes embedded in the AGN accretion disk \cite{Tagawa2023}, as well as muon pair production \cite{Hooper2023, Das2024} or beta decays of neutrons \cite{Yasuda2024, Das2024}. However, the AGN corona is particularly compelling as a neutrino production site, since in this scenario, the coronal X-ray photons attenuate the high-energy $\gamma$-rays that would otherwise be expected to accompany the ${>}\,\mathrm{TeV}$ neutrinos. 

Several studies have also investigated the diffuse neutrino emission expected from the Seyfert population as a whole in light of the NGC~1068 neutrino observations. Padovani \textit{et al.}~\cite{Padovani2024b} used a template neutrino spectrum based on the disk-corona model reported in \cite{Murase2022} and assumed a linear scaling between X-ray and neutrino luminosities. They found that the Seyfert population can account for the observed neutrino flux at energies of ${\sim}\, 1\,\text{--}\, 10\,\mathrm{TeV}$, but an additional population is needed to explain the flux at higher energies. Other works have explored the diffuse contribution using alternative models for neutrino production in Seyfert galaxies, including a leaky-box model with fixed proton injection \cite{Ambrosone2024} and a strong turbulence acceleration scenario \cite{Fiorillo2025}, finding results that slightly overshoot or are broadly consistent with the IceCube observations, respectively.

In this work, we aim to explore the constraints of the diffuse flux from the Seyfert galaxy population for the key physical parameters of the disk-corona model and quantify how unusual a source such as NGC~1068 is expected to be in this context. Furthermore, we also investigate the variance in the model parameters that is required to explain neutrino observations from the directions of other nearby Seyfert galaxies. 

We present the details of our neutrino emission model in Section~\ref{sec:nu_model}. This model is then fit to the public IceCube data for the case of NGC~1068 in Section~\ref{sec:fit}, including an evaluation of the $\gamma$-ray flux predictions. Using the result of this fit as a benchmark, we extrapolate our model to the entire population of Seyfert galaxies discussed in Section~\ref{sec:population} and calculate the resulting diffuse neutrino flux in Section~\ref{sec:results}. We discuss our results in Section~\ref{sec:discussion} and conclude in Section~\ref{sec:conclusions}. 

% -------------------------------------------------------------------------------------------------

\section{Neutrino emission model for a single source}
\label{sec:nu_model}

The corona of an AGN is a magnetized plasma of hot electrons located in the vicinity of the central region of the accretion disk \cite{Haardt1991, Miller2000, Merloni2001, Liu2002, Io2014, Jiang2014, Jiang2019}. It is very luminous in X-rays, which are produced via inverse Compton scattering of optical/UV disk photons (e.g.~\cite{Zdziarski1996, Haardt1997}). However, the exact geometry and origin of the corona are not yet fully understood. 

Similar to the model proposed by Murase \textit{et al.}~\cite{Murase2020}, we assume that inside AGN coronae, protons are accelerated via stochastic acceleration in magnetized turbulence up to energies of a few hundred TeV \cite{Lemoine2020, Lemoine2024b, Lemoine2025}. Subsequently, they undergo $pp$ interactions with ambient gas and $p\gamma$ interactions with coronal X-rays, producing high-energy neutrinos and $\gamma$-rays. The latter are attenuated in $\gamma\gamma$ interactions with X-rays, initiating electromagnetic cascades and escaping the corona at MeV energies. As in \cite{Murase2020}, we do not consider any primary electron acceleration.

We describe the relevant properties of the corona of a Seyfert galaxy as a function of its intrinsic 2--10 keV X-ray luminosity, $L_X$, following \cite{Murase2020}. Making use of the fact that the non-thermal proton spectrum resulting from stochastic acceleration follows a power law with an exponential cutoff, we then calculate the neutrino and the cascaded $\gamma$-ray spectrum of the source. To this end, we use the multi-messenger simulation code \texttt{AM\textsuperscript{3}} \cite{Klinger2024}.

\subsection{Properties of the AGN corona}
\label{sec:corona_properties}

We model the AGN corona as a sphere with radius $R_c = r_c R_S$, where $r_c$ is the dimensionless coronal radius and $R_S = 2GM_\mathrm{BH}/c^2$ is the Schwarzschild radius of the central SMBH. The typical black-hole mass of an AGN with X-ray luminosity $L_X$ is given by
\begin{equation}
M_\mathrm{BH} \simeq 2\times 10^7 M_\odot \left( \frac{L_X}{1.16 \times 10^{43} \,\mathrm{erg}/ \mathrm{s}} \right)^{0.746} \,,
\end{equation}
which is an empirical relation derived from X-ray observations of a sample of AGNs \cite{Mayers2018}.

The intrinsic X-ray spectrum of the corona is described by a single power law with an exponential cutoff,
\begin{equation}
\left( \frac{\mathrm{d}N_\gamma}{\mathrm{d}E_\gamma \mathrm{d}t} \right)_X \propto E_\gamma^{-\Gamma_X} \exp\left( -E_\gamma / E_X^\mathrm{max} \right) \,.
\label{eq:Xray_spectrum}
\end{equation}
The spectral index, $\Gamma_X$, and the cutoff energy, $E_X^\mathrm{max}$, are related to the Eddington ratio, $\lambda_\mathrm{Edd} = L_\mathrm{bol}/L_\mathrm{Edd}$, as
\begin{equation}
\Gamma_X = 0.167 \log\lambda_\mathrm{Edd} + 2.0
\end{equation}
and 
\begin{equation}
E_X^\mathrm{max} = \left[ -74\log\lambda_\mathrm{Edd} + 150 \right]\,\mathrm{keV} \,,
\end{equation}
respectively \cite{Trakhtenbrot2017, Ricci2018}. $L_\mathrm{bol}$ is the bolometric luminosity, which we calculate using the relation derived in Ref.~\cite{Hopkins2007}, and $L_\mathrm{Edd}$ is the Eddington luminosity, which is given by $L_\mathrm{Edd} \simeq 1.3\times 10^{45}\,\mathrm{erg}/\mathrm{s}\, (M_\mathrm{BH}/10^7 M_\odot)$.

We calculate the Thomson optical depth of the corona, $\tau_T$, as
\begin{equation}
\tau_T \simeq 10^{\left(2.16 - \Gamma_X\right)/1.062} \, \left( \frac{k_B T_e}{1\,\mathrm{keV}} \right)^{-0.3} \,,
\end{equation}
where $k_B T_e \simeq E_X^\mathrm{max}/2$ is the electron temperature \cite{Ricci2018}. This relation, which has been derived from simulations of coronal X-ray spectra for different values of $\tau_T$ and $k_B T_e$, yields $\tau_T\,{\sim}\,0.3\,\text{--}\,0.6$ for $42 \leq \log L_X \leq 46$ (see Table~\ref{tab:corona_properties}), i.e.~in our model, the corona is optically thin for all considered X-ray luminosities. However, other recent studies suggest that some AGNs may also have optically thick coronae with $\tau_T > 1$ \cite{Kara2017, Kamraj2022}. Assuming that the fraction of electron-positron pairs in the coronal plasma is negligibly small, we set the thermal proton number density, $n_p$, to be equal to the thermal electron number density, $n_e$, which results in 
\begin{equation}
n_p = n_e = \frac{\tau_T}{\sigma_T R_c}
\label{eq:thermal_p_density}
\end{equation}
with the Thomson cross section $\sigma_T$ (e.g.~\cite{Murase2020, Inoue2021}).

The plasma beta is defined as $\beta = P_\mathrm{th}/P_\mathrm{mag}$, where $P_\mathrm{th} = n_p k_B T_p$ is the thermal gas pressure at the virial temperature
\begin{equation}
T_p = \frac{GM_\mathrm{BH}m_p}{3R_ck_B} = \frac{m_pc^2}{6k_Br_c}
\end{equation}
and $P_\mathrm{mag} = B^2/8\pi$ is the magnetic pressure. In general, the plasma beta in AGN coronae is not very well constrained. Results of various magnetohydrodynamics simulations point towards rather low values of $\beta \lesssim 1$ \cite{Miller2000, Jiang2014, Jiang2019}. However, Inoue \textit{et al.}~\cite{Inoue2024} argue that the absence of powerful jets and the resulting lack of large-scale poloidal magnetic fields in the coronal region of Seyfert galaxies leads to $\beta\,{\sim}\, 10\,\text{--}\,100$. In line with other works studying AGN coronae as possible neutrino sources \cite{Murase2020, Eichmann2022, Das2024, Ambrosone2024, Lemoine2025}, we set $\beta = 1$. The strength of the magnetic field inside the corona can then be calculated as 
\begin{equation}
B = \sqrt{\frac{8\pi n_pk_BT_p}{\beta}} \,.
\label{eq:mag_field}
\end{equation}
The properties of the AGN corona for different X-ray luminosities and $r_c = 5$ are summarized in Table~\ref{tab:corona_properties}.

\begin{table}
\caption{Properties of the AGN corona for different $L_X$ and $r_c = 5$. $L_X$ is given in units of $\mathrm{erg}/\mathrm{s}$ and $n_p$ in units of $\mathrm{cm}^{-3}$.}
\begin{ruledtabular}
\begin{tabular}{ccccc}
$\log L_X$ & $\log M_\mathrm{BH}/M_\odot$ & $\tau_T$ & $\log n_p$ & $B\, [\mathrm{kG}]$ \\
\hline
42.0 & 6.51 & 0.56 & 11.25 & 15.0 \\
43.0 & 7.25 & 0.49 & 10.45 & 5.9 \\
44.0 & 8.00 & 0.43 & 9.64 & 2.3 \\
45.0 & 8.74 & 0.37 & 8.84 & 0.93 \\
46.0 & 9.49 & 0.34 & 8.05 & 0.37 \\
\end{tabular}
\end{ruledtabular}
\label{tab:corona_properties}
\end{table}

\subsection{Proton acceleration}
\label{sec:acceleration}

The presence of strong and turbulent magnetic fields inside AGN coronae suggests that they are possible sites of stochastic particle acceleration in magnetized turbulence, also referred to as second-order Fermi acceleration. Protons are accelerated via repeated scattering by plasma waves, causing them to gradually gain energy. For this process to be efficient, not only a strong magnetic field, but also a sufficient level of turbulence is required. We quantify the latter in terms of the inverse turbulence strength, $\eta$, defined as
\begin{equation}
\eta = \frac{B^2}{\delta B^2} = \frac{B_0^2 + \delta B^2}{\delta B^2} \,,
\end{equation} 
where $B_0$ and $\delta B$ are the regular and the turbulent component of the magnetic field, respectively.

Stochastic acceleration can be modeled as diffusion in momentum space described by a momentum diffusion coefficient. For relativistic particles, this diffusion coefficient can be approximated as
\begin{equation}
D_{E_p} \simeq \frac{c}{\eta R_c} \left( \frac{V_A}{c} \right)^2 \left( \frac{E_p}{eBR_c} \right)^{q-2} E_p^2 \,,
\label{eq:diff_coeff}
\end{equation}
where $V_A = B/\sqrt{4\pi m_pn_p}$ is the Alfv\'en velocity and $q$ is the spectral index of the turbulence power spectrum \cite{Kimura2019}. For proton energies $E_p \ll E_p^\mathrm{max}$, where $E_p^\mathrm{max}$ is the energy at which cooling and escape losses start to dominate over acceleration, we approximate the resulting non-thermal proton spectrum by a power law with spectral index $1-q$ \cite{Becker2006, Stawarz2008}. Starting from the momentum diffusion equation that describes the evolution of the particle distribution function, $f$, under stochastic acceleration,
\begin{equation}
\frac{\partial f}{\partial t} = \frac{1}{E_p^2} \frac{\partial}{\partial E_p} \left[ E_p^2 D_{E_p} \frac{\partial f}{\partial E_p} \right] \,,
\end{equation}
it can be seen that in a stationary system ($\partial_t f = 0$), $f$ has to satisfy $E_p^2 D_{E_p} \partial_{E_p}f = \mathrm{const.}$ With $D_{E_p} \propto E_p^q$ (cf.~Eq.~(\ref{eq:diff_coeff})) and  $\mathrm{d}n_p/\mathrm{d}E_p = 4\pi E_p^2 f/c^3$ this yields $\mathrm{d}n_p/\mathrm{d}E_p \propto E_p^{1-q}$. Thus, for Kolmogorov turbulence with $q = 5/3$, the resulting proton spectrum is a power law with spectral index $-2/3$ and an exponential cutoff at $E_p^\mathrm{max}$,
\begin{equation}
\frac{\mathrm{d}N_p}{\mathrm{d}E_p} \propto E_p^{-2/3} \exp\left( -E_p/E_p^\mathrm{max} \right) \,.
\label{eq:proton_spectrum}
\end{equation}
The corresponding acceleration timescale is given by (e.g.~\cite{Murase2020, Kimura2019})
\begin{equation}
t_\mathrm{acc} \simeq \frac{E_p^2}{D_{E_p}} = \eta \left( \frac{c}{V_A} \right)^2 \frac{R_c}{c} \left( \frac{E_p}{eBR_c} \right)^{2-q} \,.
\label{eq:t_acc}
\end{equation}

\begin{table}[t]
\caption{Proton luminosities corresponding to a pressure ratio of $P_\mathrm{CR}/P_\mathrm{th} = 0.5$ and their ratio to the Eddington luminosity for different $L_X$ and $\eta$. All luminosities are calculated for a coronal radius of $r_c = 5$.}
\begin{ruledtabular}
\begin{tabular}{ccccc}
 & \multicolumn{2}{c}{$\eta = 1$} & \multicolumn{2}{c}{$\eta = 100$} \\
 \cline{2-3}\cline{4-5}
$L_X\, [\mathrm{erg}/\mathrm{s}]$ & $L_p\,[\mathrm{erg}/\mathrm{s}]$ & $L_p/L_\mathrm{Edd}$ & $L_p\,[\mathrm{erg}/\mathrm{s}]$ & $L_p/L_\mathrm{Edd}$ \\\hline
$10^{42}$ & $1.6\times 10^{43}$ & 0.039 & $4.0\times 10^{42}$ & 0.010 \\
$10^{44}$ & $4.9\times 10^{44}$ & 0.038 & $1.0\times 10^{44}$ & 0.008 \\
$10^{46}$ & $1.3\times 10^{46}$ & 0.031 & $2.7\times 10^{45}$ & 0.007 \\
\end{tabular}
\end{ruledtabular}
\label{tab:proton_luminosities}
\end{table}

We determine the normalization of the proton spectrum in Eq.~(\ref{eq:proton_spectrum}) by the total energy injected into proton acceleration, which can be expressed in terms of the ratio of the cosmic-ray (CR) pressure to the thermal gas pressure inside the corona, $P_\mathrm{CR}/P_\mathrm{th}$, with
\begin{equation}
P_\mathrm{CR} = \frac{1}{3}\int\mathrm{d}E_p\, E_p \frac{\mathrm{d}N_p}{\mathrm{d}E_p\mathrm{d}V} \,.
\end{equation}
We set $P_\mathrm{CR}/P_\mathrm{th} \leq 0.5$ as a strict upper limit on this pressure ratio \cite{Kheirandish2021}. The reasoning behind this is as follows: In the limit $P_\mathrm{CR}/P_\mathrm{th} = 0.5$, the total kinetic energy of non-thermal particles is equal to half of the gravitational binding energy of the corona. According to the virial theorem, in this case all the kinetic energy of the system is contained in the non-thermal particle population and there is no room for any kinetic energy in the thermal particle population. This means that for $P_\mathrm{CR}/P_\mathrm{th} > 0.5$, the corona is no longer stable. The injected CR proton luminosities that correspond to a pressure ratio of $P_\mathrm{CR}/P_\mathrm{th} = 0.5$ are listed in Table \ref{tab:proton_luminosities} for different $L_X$ and two benchmark values of $\eta$. For all considered X-ray luminosities and turbulence levels, the required proton luminosity is much lower than the Eddington luminosity, with a maximum of $L_p/L_\mathrm{Edd} \,{\sim}\, 4\%$ for $\eta = 1$ and $L_X = 10^{42}\,\mathrm{erg}/\mathrm{s}$.

\subsection{Proton cooling and escape}
\label{sec:losses}

Relevant cooling processes for non-thermal protons inside AGN coronae include $pp$ interactions, $p\gamma$ interactions, Bethe-Heitler pair production, and synchrotron radiation \cite{Murase2020, Eichmann2022}. 

The target photon spectrum for $p\gamma$ interactions and Bethe-Heitler pair production is the sum of the optical/UV spectrum of the accretion disk and the X-ray spectrum of the corona. The disk spectrum is a multi-temperature blackbody spectrum, which we approximate as
\begin{equation}
\left( \frac{\mathrm{d}N_\gamma}{\mathrm{d}E_\gamma\mathrm{d}t} \right)_\mathrm{disk} \propto E_\gamma^{-5/3} \exp\left( -E_\gamma/k_B T_\mathrm{max} \right)
\label{eq:disk_spectrum}
\end{equation}
for photon energies $E_\gamma \geq k_B T(R_\mathrm{out})$ \cite{Ghisellini2013}. The maximum temperature of the disk near the SMBH, $T_\mathrm{max}$, is given by 
\begin{equation}
T_\mathrm{max} \simeq \left(\frac{L_\mathrm{disk}}{200\pi R_S^2 \sigma_\mathrm{SB}}\right)^{1/4} \,,
\end{equation}
where $L_\mathrm{disk} \simeq 0.5 L_\mathrm{bol}$ \cite{Murase2020} is the disk luminosity and $\sigma_\mathrm{SB}$ is the Stefan-Boltzmann constant. $T(R_\mathrm{out})$ denotes the temperature at the outer radius of the accretion disk, which is $k_B T(R_\mathrm{out}) \simeq 0.2\,\mathrm{eV}$ for $R_\mathrm{out} = 10^3 R_S$ (cf.~Eq.~(8.5) in Ref.~\cite{Ghisellini2013}). As already explained in Section~\ref{sec:corona_properties}, we describe the X-ray spectrum of the corona by a single power law with an exponential cutoff. The resulting disk-corona spectrum, which is obtained by summing Eqs.~(\ref{eq:Xray_spectrum}) and (\ref{eq:disk_spectrum}), then only depends on the X-ray luminosity of the source.

The proton energy threshold for photopion production is $E_{p\text{--}\pi}^\mathrm{th} \simeq 3.4\,\mathrm{PeV}\, (E_\gamma/10\,\mathrm{eV})^{-1}$ \cite{Sigl2017}. Thus, the proton energies that we consider are not high enough for photopion production with disk photons and the main target photons for $p\gamma$ interactions are X-rays emitted by the corona. The optical/UV disk spectrum is mainly relevant for Bethe-Heitler pair production with a threshold energy of $E_{p\text{--}\mathrm{BH}}^\mathrm{th} \simeq 48\,\mathrm{TeV}\, (E_\gamma/10\,\mathrm{eV})^{-1}$ \cite{Sigl2017}, i.e.~the required proton energy is approximately two orders of magnitude lower than for $p\gamma$ interactions.

Besides the cooling processes mentioned above, we also take into account particle escape via infall onto the SMBH and diffusion. The infall timescale is given by 
\begin{equation}
t_\mathrm{fall} = \frac{R_c}{\alpha V_K} 
\end{equation}
with the viscosity parameter, $\alpha = 0.1$, and the Keplerian velocity, $V_K = \sqrt{GM_\mathrm{BH}/R_c} = c/\sqrt{2r_c}$ \cite{Murase2020}. The timescale for diffusive escape from the corona can be estimated as (e.g.~\cite{Stawarz2008, Murase2020})
\begin{equation}
t_\mathrm{diff} \simeq \frac{9}{\eta} \frac{R_c}{c} \left( \frac{E_p}{eBR_c} \right)^{q-2} \,.
\end{equation}
We calculate the cutoff energy of the proton spectrum, $E_p^\mathrm{max}$, by solving $t_\mathrm{acc} (E_p^\mathrm{max}) = t_\mathrm{loss}(E_p^\mathrm{max})$ with the acceleration timescale given in Eq.~(\ref{eq:t_acc}) and the total loss timescale for protons given by
\begin{equation}
t_\mathrm{loss}^{-1} = t_{pp}^{-1} + t_{p\gamma}^{-1} + t_\mathrm{BH}^{-1} + t_\mathrm{syn}^{-1} + t_\mathrm{diff}^{-1} + t_\mathrm{fall}^{-1}\,.
\label{eq:t_loss}
\end{equation}
The timescales for $pp$ interactions, $p\gamma$ interactions, Bethe-Heitler (BH) pair production, and synchrotron radiation are calculated directly by the multi-messenger simulation code \texttt{AM\textsuperscript{3}} \cite{Klinger2024}, which we use to compute the resulting neutrino and $\gamma$-ray emission. The relative importance of the different cooling and escape processes at different X-ray luminosities is discussed in Appendix \ref{sec:proton_timescales}.

\subsection{Neutrino spectra}

Inside the AGN corona, high-energy neutrinos can be produced in both $pp$ interactions with ambient gas and $p\gamma$ interactions with X-rays. Taking into account all energy and escape loss processes for protons described in Section~\ref{sec:losses}, we calculate the neutrino spectrum of a source with a given X-ray luminosity using \texttt{AM\textsuperscript{3}}, an open-source code that allows to simulate the non-thermal particle emission of an astrophysical source \cite{Klinger2024}. As mentioned in Section \ref{sec:corona_properties}, the coronal radius is taken to be a multiple of the Schwarzschild radius, $R_c = r_c R_S$. The density of thermal protons, which act as a target for $pp$ interactions, is given by Eq.~(\ref{eq:thermal_p_density}), while the magnetic field strength is set according to Eq.~(\ref{eq:mag_field}). The injected non-thermal proton spectrum as well as the target photon field relevant for $p\gamma$ and Bethe-Heitler interactions are specified in Sections \ref{sec:acceleration} and \ref{sec:losses}, respectively.

\begin{figure}[t]
\includegraphics[width=8.5cm]{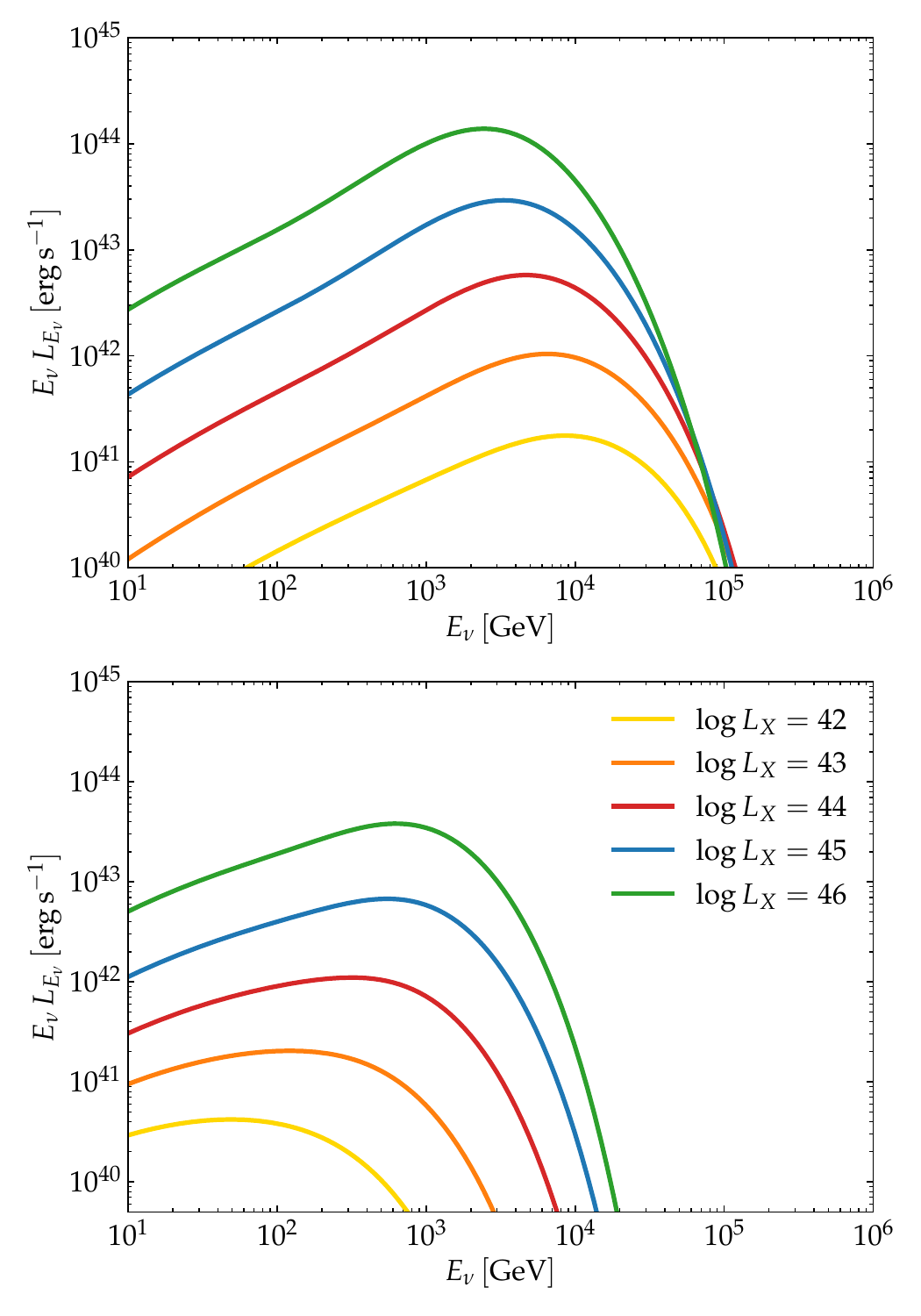}
\caption{Neutrino spectra for different X-ray luminosities computed for $\eta = 10$ (top panel) and $\eta = 150$ (bottom panel). The neutrino spectra are normalized so that $P_\mathrm{CR}/P_\mathrm{th} = 0.5$ and the dimensionless coronal radius is set to $r_c = 5$. $E_\nu L_{E_\nu} \equiv E_\nu^2 \,\mathrm{d}N_\nu/\mathrm{d}E_\nu\mathrm{d}t$ denotes the differential all-flavor neutrino luminosity.}
\label{fig:neutrino_spectra}
\end{figure}

Fig.~\ref{fig:neutrino_spectra} shows the resulting neutrino spectra for different X-ray luminosities and two different values of $\eta$. In general, sources with a higher X-ray luminosity also have a higher neutrino luminosity. Given the assumption that the pressure ratio $P_\mathrm{CR}/P_\mathrm{th}$ is constant across sources of all X-ray luminosities, we find that the total neutrino luminosity, $L_\nu$, roughly  follows a power-law scaling of the form
\begin{equation}
L_\nu \propto L_X^{0.7}\,.
\label{eq:Lnu_Lx_scaling}
\end{equation}
This means that, in our model, the neutrino luminosity increases slightly less than linearly with the X-ray luminosity. The turbulence strength parameter, $\eta$, determines the efficiency of the proton acceleration process and thus the cutoff energy of the proton spectrum and also of the resulting neutrino spectrum. Smaller values of $\eta$ corresponding to very turbulent magnetic fields inside the corona lead to higher cutoff energies since proton acceleration is more efficient. In this case, the acceleration process is stopped by $p\gamma$ and Bethe-Heitler interactions, and we find lower cutoff energies for sources with higher $L_X$ due to the higher target photon density inside the corona. For larger values of $\eta$, the confinement time of protons inside the corona becomes shorter and the acceleration process is stopped by protons escaping the corona via diffusion. This leads to higher cutoff energies for sources with higher $L_X$ since more luminous sources have larger coronae. 

Overall, both $pp$ and $p\gamma$ interactions contribute to the production of neutrinos. Only for low X-ray luminosities in combination with large $\eta$, the contribution from $p\gamma$ interactions becomes subdominant compared to that from $pp$ interactions due to the low cutoff energy of the corresponding proton spectrum. A more detailed discussion of the relative importance of the different cooling and escape processes for protons can be found in Appendix \ref{sec:proton_timescales}.

\subsection{Cascaded $\gamma$-ray spectra}
\label{sec:gamma_ray_spectra}

In addition to neutrinos, hadronic interactions also produce high-energy $\gamma$-rays with an intrinsic luminosity comparable to that of neutrinos. Unlike neutrinos, however, these $\gamma$-rays do not simply escape from the corona. Instead, they are attenuated in $\gamma\gamma$ interactions with X-rays, which initiate electromagnetic cascades inside the corona \cite{Murase2020, Murase2022, Eichmann2022, Das2024}. Secondary electrons and positrons produced in $pp$, $p\gamma$ and Bethe-Heitler interactions also contribute to these electromagnetic cascades through both synchrotron radiation and inverse Compton scattering.

\begin{figure}[t]
\includegraphics[width=8.5cm]{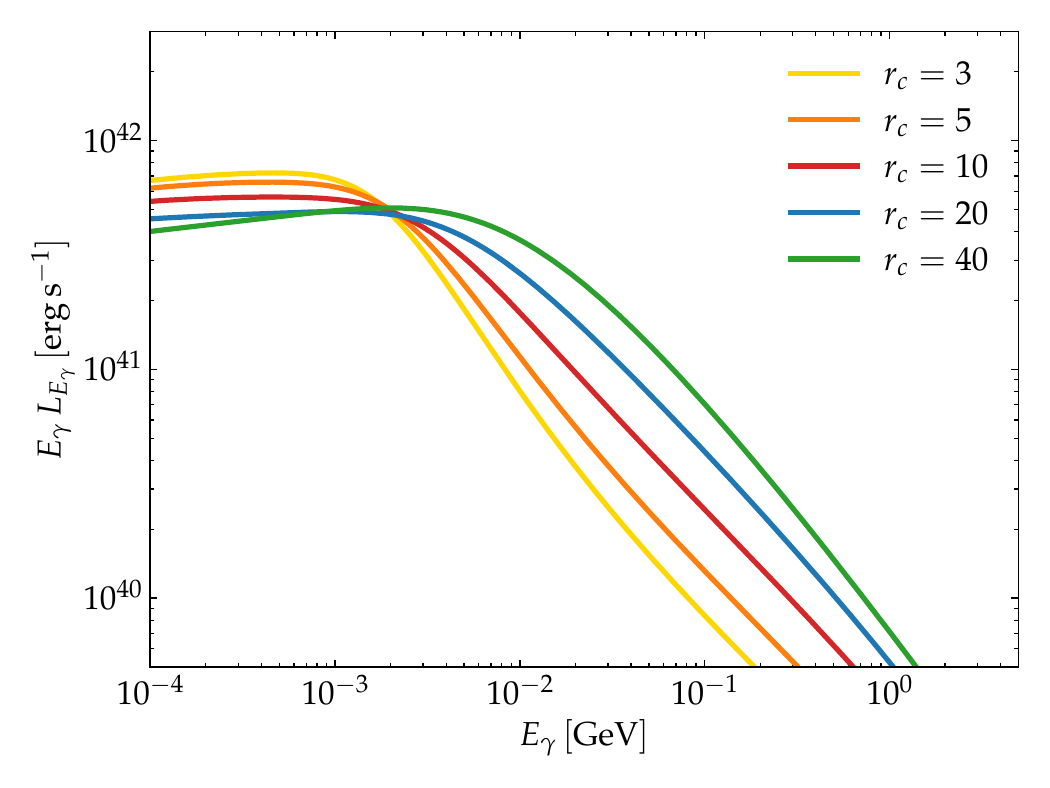}
\caption{Cascaded $\gamma$-ray spectra for different emission radii calculated for the X-ray luminosity of NGC 1068 ($L_X = 4.17\times 10^{43}\,\mathrm{erg}/\mathrm{s}$) and $\eta = 20$. Here, the underlying proton spectra are normalized so that $L_p/L_X = 0.5$. $E_\gamma\, L_{E_\gamma} \equiv E_\gamma^2 \,\mathrm{d}N_\gamma/\mathrm{d}E_\gamma\mathrm{d}t$ denotes the differential $\gamma$-ray luminosity.}
\label{fig:cascaded_spectra}
\end{figure}

The interaction timescales for the leptonic processes mentioned above are much shorter than those for the hadronic processes discussed in Section \ref{sec:losses}. A single \texttt{AM\textsuperscript{3}} simulation that self-consistently accounts for both hadronic and leptonic interactions would therefore require a solver time step small enough to resolve all leptonic processes as well as a sufficiently long total simulation time for the hadronic component to reach a steady state. This would make the computation very time-consuming and thus unfeasible across a grid of source parameters. To address this, we decouple the purely leptonic part of the simulation, which describes the electromagnetic cascade, from the hadronic part and perform two consecutive simulations. First, we run a purely hadronic \texttt{AM\textsuperscript{3}} simulation to compute the neutrino spectrum as well as the intrinsic $\gamma$-ray and electron-positron spectra. Then, we inject the latter together with the photon spectrum of the accretion disk and the corona into a second, purely leptonic \texttt{AM\textsuperscript{3}} simulation in order to calculate the spectrum of cascaded $\gamma$-rays escaping from the corona of the source.

As shown in Fig.~\ref{fig:cascaded_spectra}, only $\gamma$-rays with energies $E_\gamma \lesssim 10\,\mathrm{MeV}$ can escape the corona completely unhindered. The shape and, in particular, the cutoff energy of the cascaded $\gamma$-ray spectrum depends strongly on the emission radius. Smaller values of $r_c$ lead to lower cutoff energies and less $\gamma$-rays at energies above $\sim\,$10$\,\mathrm{MeV}$ due to the higher density of target photons for $\gamma\gamma$ interactions at smaller emission radii. In addition to $\eta$ and the total energy injected into proton acceleration, $r_c$ is therefore an important model parameter that shapes the multi-messenger spectrum of a source with a given X-ray luminosity.

% -------------------------------------------------------------------------------------------------

\section{Model fit for NGC 1068}
\label{sec:fit}

In order to constrain the remaining free parameters of the neutrino emission model presented in Section~\ref{sec:nu_model}, we fit the neutrino spectrum for NGC 1068 to publicly available neutrino data. Moreover, we compare the resulting cascaded $\gamma$-ray emission to existing upper limits in the MeV-GeV range to ensure that our model predictions are consistent with both neutrino and $\gamma$-ray observations of NGC 1068.

\subsection{Fit of the neutrino spectrum}
\label{sec:fit_nu_spectrum}

Besides the coronal radius, $r_c$, the only free parameters of our model are the inverse turbulence strength of the coronal magnetic field, $\eta$, and the total energy injected into proton acceleration, expressed in terms of the pressure ratio, $P_\mathrm{CR}/P_\mathrm{th}$. The neutrino spectrum of a source is primarily determined by the latter two parameters. To constrain them, we fit our model neutrino spectrum for NGC 1068 to the public 10-year IceCube dataset of track-like events for point-source searches \cite{IceCube2021}, which consists of a list of muon events with reconstructed energies, $\hat{E}$, and directions, $\hat{\theta}$. We do this by performing a fit with an unbinned likelihood function using the open-source software \texttt{SkyLLH} \cite{Braun2008, Wolf2019, Bellenghi2023}. The likelihood function
\begin{equation}
\mathcal{L}(n_s, \eta) = \prod_{i=1}^N \left[ \frac{n_s}{N}\, \mathcal{S} ( \hat{E}_i, \hat{\theta}_i|\eta ) + \left( 1 - \frac{n_s}{N}\right) \mathcal{B}(\hat{E}_i, \hat{\theta}_i) \right]
\label{eq:likelihood}
\end{equation}
is a weighted sum of the signal probability density $\mathcal{S}(\hat{E}_i, \hat{\theta}_i|\eta)$, which gives the probability that the $i$th event is a signal event associated with the source, and the background probability density $\mathcal{B}(\hat{E}_i, \hat{\theta}_i)$, which gives the probability that the $i$th event is a background event. Here, $n_s$ denotes the number of signal events and $N$ is the total number of events in the dataset. To obtain the values of $\eta$ and $n_s$ for which our model for NGC 1068 fits the data best, we first compute the neutrino spectra for the X-ray luminosity of NGC 1068 for $\eta = 1, 2, 3, \dots, 150$. Then, we determine the values of $\eta$ and $n_s$ that maximize the likelihood given in Eq.~(\ref{eq:likelihood}). In doing so, we impose bounds on $n_s$ to ensure that $P_\mathrm{CR}/P_\mathrm{th} \leq 0.5$ is satisfied at all times. 

\begin{figure}[t]
\includegraphics[width=8.5cm]{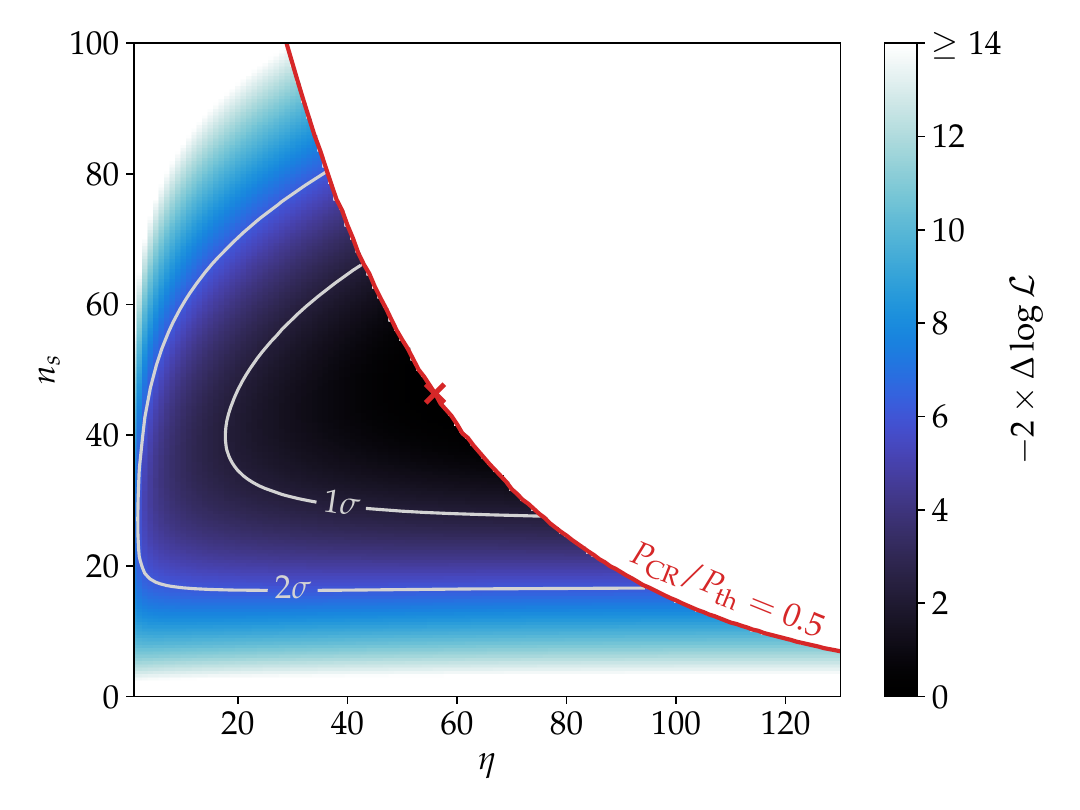}
\caption{Log-likelihood ratio as a function of $n_s$ and $\eta$. The best-fit parameters obtained for NGC 1068 are indicated by a red cross and the gray contours represent the $1\sigma$ and $2\sigma$ confidence regions according to Wilks' theorem \cite{Wilks1938}. The region above the red line corresponds to $P_\mathrm{CR}/P_\mathrm{th} > 0.5$ and is thus excluded.}
\label{fig:contour_plot_NGC1068}
\end{figure}

For a dimensionless coronal radius of $r_c = 5$, our fit yields $\hat{n}_s = 46.4$ signal events and an inverse turbulence strength of $\hat{\eta} = 56$, which corresponds to a total proton luminosity of $L_p = 1.43L_X \simeq 6\times 10^{43}\,\mathrm{erg}/\mathrm{s} \simeq 0.009 L_\mathrm{Edd}$ and a pressure ratio of $P_\mathrm{CR}/P_\mathrm{th} = 0.5$. This is consistent with Refs.~\cite{Murase2020, Kheirandish2021, Eichmann2022, Blanco2023, Das2024}, all of which report that $L_p \simeq L_X$ and $P_\mathrm{CR}/P_\mathrm{th} \simeq 0.5$ is necessary to explain the observed neutrino flux of NGC 1068. Fig.~\ref{fig:contour_plot_NGC1068} shows the log-likelihood ratio 
\begin{equation}
-2 \times \Delta \log \mathcal{L} \equiv -2\log \frac{\mathcal{L}(n_s, \eta)}{\mathcal{L}(\hat{n}_s, \hat{\eta})}
\end{equation}
as a function of the parameters $n_s$ and $\eta$. By setting an upper limit on $P_\mathrm{CR}/P_\mathrm{th}$, we restrict the allowed values of $n_s$ and $\eta$ to only part of the considered parameter space. However, when performing the same fit without imposing a constraint on the pressure ratio, we find that our result obtained with $P_\mathrm{CR}/P_\mathrm{th} \leq 0.5$ still lies within the $1\sigma$ confidence region of the fit result obtained without an upper limit on $P_\mathrm{CR}/P_\mathrm{th}$. This indicates that requiring $P_\mathrm{CR}/P_\mathrm{th} \leq 0.5$ does not result in a significantly worse fit to the data.

\begin{figure}[t]
\includegraphics[width=8.5cm]{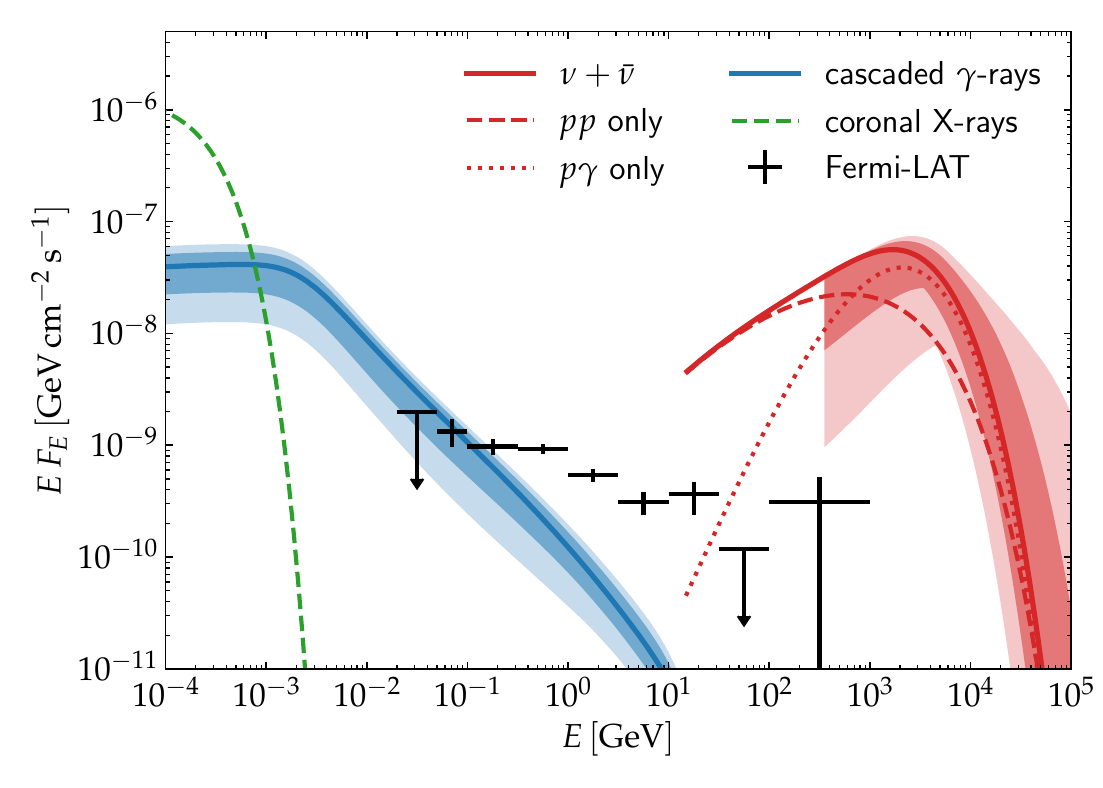}
\caption{Best-fit all-flavor neutrino spectrum of NGC 1068 (red) together with the corresponding cascaded $\gamma$-ray spectrum (blue). The shaded bands indicate the $1\sigma$ and $2\sigma$ uncertainty regions of the flux and, for the neutrino spectrum, the central 68\% energy range. The latest Fermi-LAT data of NGC 1068 \cite{Ajello2023} is shown for comparison.}
\label{fig:mm_spectrum_NGC1068}
\end{figure}

Our best-fit model neutrino spectrum for NGC 1068 is shown in Fig.~\ref{fig:mm_spectrum_NGC1068}, together with the $1\sigma$ and $2\sigma$ uncertainty bands. The latter correspond to the confidence regions marked in Fig.~\ref{fig:contour_plot_NGC1068} and are calculated assuming Wilks' theorem \cite{Wilks1938} as implemented in \texttt{SkyLLH}. The asymmetrical shape of the uncertainty bands is due to the fact that the best-fit values for $n_s$ and $\eta$ lie at the boundary of the parameter space. We find that both $pp$ and $p\gamma$ interactions are important for the production of neutrinos. Around the peak of the spectrum at $\sim\,$1.5$\,\mathrm{TeV}$, most of the neutrino emission is produced in $p\gamma$ interactions, while at lower energies the spectrum is dominated by neutrinos from $pp$ interactions. The sensitive energy range of our analysis, defined as the central 68\% interval of neutrino energies contributing to the excess over the background \cite{Schoenen2017, Bellenghi2024, Kontrimas2025}, extends from $350\,\mathrm{GeV}$ to $190\,\mathrm{TeV}$. A more detailed discussion of the method used to determine this energy range, as well as a comparison of our result to the best-fit power law spectrum of NGC 1068, can be found in Appendix \ref{sec:sensitive_energy_range}.

Similar to Ref.~\cite{Capel2024}, we use posterior predictive checks to evaluate the goodness of fit. Using \texttt{SkyLLH}, we simulate 10000 pseudo datasets assuming our best-fit model neutrino spectrum and select all events in a circular region with radius $r=5^\circ$ around the position of NGC 1068. In 24 logarithmically spaced energy bins between $100\,\mathrm{GeV}$ and $1\,\mathrm{PeV}$, we quantify the discrepancy between the simulated and the observed data and find that it is smaller than $2\sigma$ in all bins. This indicates a good agreement between our model and the data.

\subsection{Cascaded $\gamma$-ray spectrum of NGC 1068}
\label{sec:gamma_ray_spectrum_NGC1068}

To ensure that our model predictions do not violate $\gamma$-ray observations, we also calculate the cascaded $\gamma$-ray spectrum that corresponds to the best-fit neutrino spectrum of NGC 1068. The resulting spectrum is shown in Fig.~\ref{fig:mm_spectrum_NGC1068} together with the $1\sigma$ and $2\sigma$ uncertainty bands calculated from the uncertainty bands around the best-fit neutrino spectrum. Comparing our result to the latest Fermi-LAT observations of NGC 1068 \cite{Ajello2023}, we find that for a dimensionless coronal radius of $r_c = 5$, our predicted $\gamma$-ray spectrum agrees well with the observational data. Especially the $95\%$ upper limit at $\sim\,$30$\,\mathrm{MeV}$, which provides the strongest constraint, lies within the $1\sigma$ uncertainty band.

This is in agreement with the results obtained by Das \textit{et al.}~\cite{Das2024}, who find $R_c \lesssim (3\,\text{--}\,15) R_S$ considering a purely photohadronic scenario. A similar study using a previously reported $\gamma$-ray spectrum of NGC 1068 that extends only down to $\sim\,$70$\,\mathrm{MeV}$ \cite{Abdollahi2020} yields somewhat looser constraints on the coronal radius with $R_c \lesssim 30 R_S$ for a $p\gamma$-only scenario and $R_c \lesssim 100 R_S$ for a $pp$-only scenario \cite{Murase2022}. In addition, various indirect observational methods that are used to determine the size of AGN coronae also point towards rather small and compact coronae with a typical radius of $R_c \lesssim 5R_S$ \cite{Laha2025}. The fact that the observed $\gamma$-ray spectrum of NGC 1068 at energies $\gtrsim 500\,\mathrm{MeV}$ cannot be explained by hadronic interactions within the corona is consistent with the results of multi-messenger studies \cite{Eichmann2022, Ajello2023, Salvatore2024}, which suggest that these $\gamma$-rays are mainly produced in the starburst region. It is expected that the starburst only contributes significantly at energies above ${\sim}\, 100\,\mathrm{MeV}$ \cite{Ajello2023}. However, if there was another additional source of $\gamma$-rays with energies ${\lesssim}\,100\,\mathrm{MeV}$, e.g.~due to primary electron acceleration, an even smaller coronal radius or non-uniform $\gamma$-ray absorption in the corona would be necessary in order not to exceed the Fermi-LAT observations \cite{Inoue2020}.

% -------------------------------------------------------------------------------------------------

\section{Source population}
\label{sec:population}

\begin{figure*}[t]
\includegraphics[width=17.5cm]{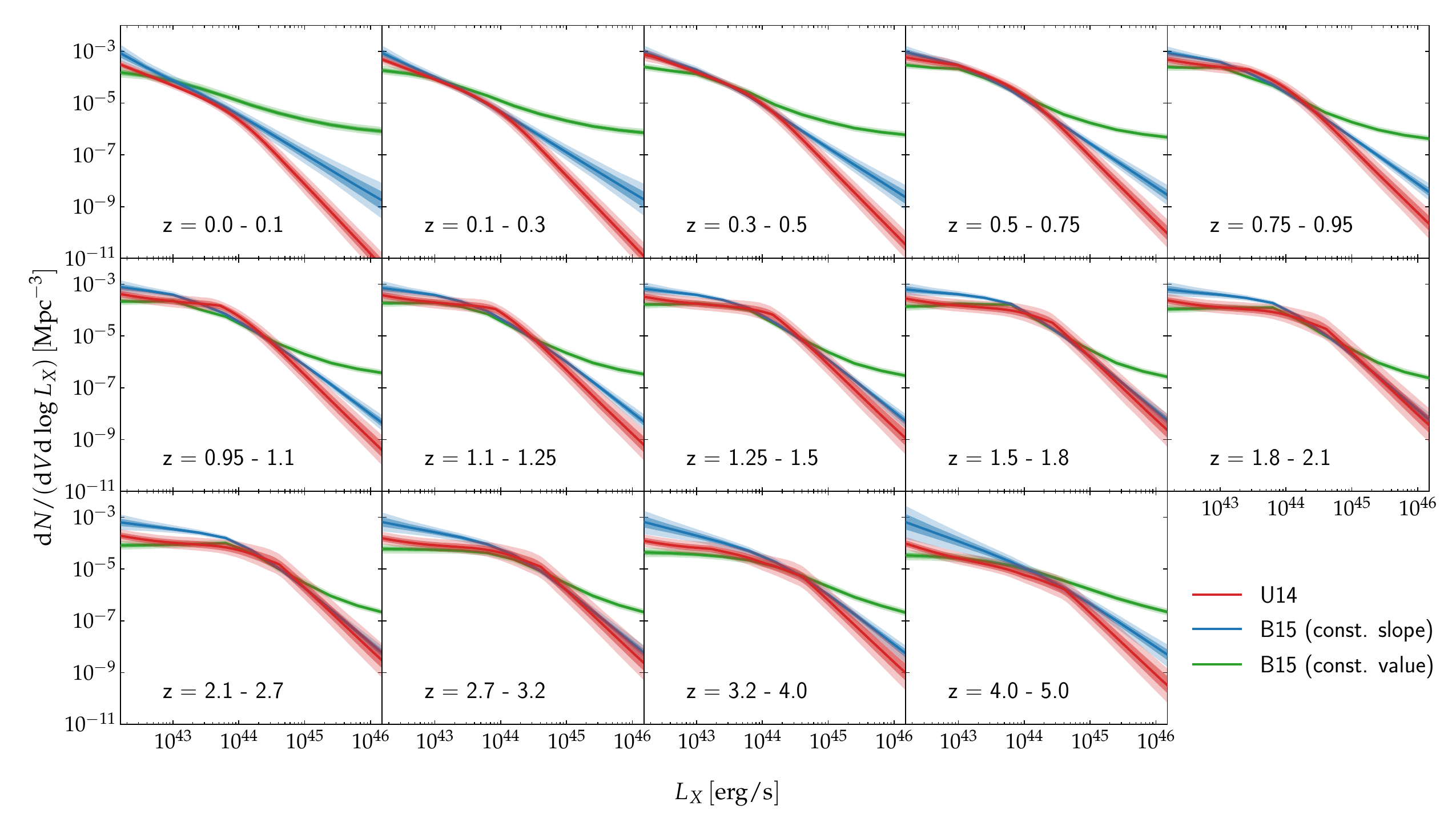}
\caption{2--10 keV X-ray luminosity functions of AGNs from Ref.~\cite{Ueda2014} (U14) and Ref.~\cite{Buchner2015} (B15). The red shaded bands correspond to the $1\sigma$ and $2\sigma$ uncertainty regions of the U14 XLF derived from the statistical errors of the model parameters assuming that all parameters are uncorrelated, i.e.~these bands indicate upper limits on the size of the actual uncertainty regions. The blue and green shaded bands indicate the 68\% and 95\% credible regions of the B15 XLFs for the constant-slope and constant-value prior, respectively.}
\label{fig:all_LFs}
\end{figure*}

We apply the neutrino emission model described in Section~\ref{sec:nu_model} to an entire population of AGNs, taking into account the fit result for NGC 1068, with the aim of calculating the resulting diffuse neutrino flux. The cosmological evolution of this population is described by the X-ray luminosity function (XLF), which gives the differential number density of AGNs per comoving volume as a function of X-ray luminosity and redshift. Here, we consider three different 2--10 keV XLFs. 

The luminosity function by Ueda \textit{et al.}~presented in Ref.~\cite{Ueda2014} (hereafter U14) is derived assuming that the XLF in the local Universe can be modeled by a smoothly broken power law, which is then multiplied by a luminosity-dependent evolution factor to obtain the XLF at a given redshift $z > 0$. The free parameters of this function are determined by a fit to the source sample. Heavily obscured Compton-thick AGNs with column densities $N_H \geq 1/\sigma_T \simeq 10^{24}\,\mathrm{cm}^{-2}$ are assumed to follow the same cosmological evolution as obscured Compton-thin AGNs with $10^{22}\,\mathrm{cm}^{-2} \leq N_H \leq 10^{24}\,\mathrm{cm}^{-2}$, which is in agreement with observations of the cosmic X-ray background. 

In contrast, Buchner \textit{et al.}~\cite{Buchner2015} (hereafter B15) use a non-parametric approach to derive the XLF. They fit a 3D histogram in $L_X$, $z$ and $N_H$ to their source sample with the only assumption being that the XLF does not vary rapidly between neighboring bins. To ensure this, two different smoothness priors are considered: the constant-slope prior and the constant-value prior. The former keeps power-law slopes intact, while the latter retains the current value of the XLF unless constraints are imposed by the data. Moreover, B15 constrain the contribution from Compton-thick AGNs in a more precise way. For each source in their sample, they fit a spectral model to determine the obscuring column density and the intrinsic X-ray luminosity consistently taking into account different sources of uncertainties.
In this way, they do not have to rely on a fit to the cosmic X-ray background, which is rather insensitive to the exact fraction of Compton-thick AGNs \cite{Akylas2012}. Although Compton-thick sources do not appear very luminous due to their high level of obscuration, they are just as important as less obscured AGNs, since in our model the neutrino emission properties of a source do not depend on $N_H$ and, unlike X-rays, neutrinos are not absorbed by the dusty torus surrounding the central region of an AGN.

\begin{table*}[t]
\caption{Properties of the source populations simulated based on the three different XLFs and the sources selected from the BASS catalog. Given are the number of sources with an intrinsic X-ray flux higher than $10^{-11}\,\mathrm{erg}\,\mathrm{cm}^{-2}\,\mathrm{s}^{-1}$ (column 2) and higher than that of NGC 1068 (column 3), the total number of simulated sources in the population (column 4), and the total intrinsic X-ray flux from the population after replacing all sources with $F_X \geq 10^{-11}\,\mathrm{erg}\,\mathrm{cm}^{-2}\,\mathrm{s}^{-1}$ by sources from the BASS catalog (column 5).}
\begin{ruledtabular}
\begin{tabular}{lcccc}
 & \# sources with & \# sources with & \# all sources & $F_X^\mathrm{tot}$ \\
 & $F_X \geq 10^{-11}\,\mathrm{erg}\,\mathrm{cm}^{-2}\,\mathrm{s}^{-1}$ & $F_X \geq F_X^\mathrm{NGC 1068}$\footnote{Here, we assume an intrinsic X-ray flux of $F_X^\mathrm{NGC1068} = 2.80\times 10^{-9}\,\mathrm{erg}\,\mathrm{cm}^{-2}\,\mathrm{s}^{-1}$ for NGC 1068, which corresponds to $L_X = 4.17\times 10^{43}\,\mathrm{erg}/\mathrm{s}$ and $d_L = 11.14\,\mathrm{Mpc}$ \cite{Marinucci2016, Tikhonov2021}.} & & $[\mathrm{erg}\,\mathrm{cm}^{-2}\,\mathrm{s}^{-1}\,\mathrm{sr}^{-1}]$ \\
\hline
U14 & 363 & 0 & $5.1 \times 10^8$ & $7.9\times 10^{-8}$ \\
B15 const.~slope & 865 & 0 & $1.3 \times 10^9$ & $9.1\times 10^{-8}$ \\
B15 const.~value & 28471 & 46 & $3.4 \times 10^8$ & $1.2\times 10^{-7}$ \\
BASS catalog & 212 & 1 & - & - \\
\end{tabular}
\end{ruledtabular}
\label{tab:source_pops}
\end{table*}

Fig.~\ref{fig:all_LFs} shows the three considered XLFs in different redshift bins between $z = 0$ and $5$ together with their uncertainties. All three XLFs have roughly the shape of a broken power law with significantly more low-luminosity sources than high-luminosity sources. However, they do not agree within their uncertainties. While the U14 XLF and the B15 XLF for the constant-slope prior have an overall similar shape, especially at intermediate redshifts, the B15 XLF for the constant-value prior is much flatter and thus predicts many more high-luminosity sources with $L_X \gtrsim 10^{45}\,\mathrm{erg}/\mathrm{s}$. The discrepancy between the two XLFs by B15, particularly at the high-luminosity end, is due to the fact that the underlying X-ray source sample consists mainly of data from deep X-ray surveys with small survey areas. These surveys are not suited to accurately determine the number density of high-luminosity AGNs, since they are very rare. Thus, the shape of the XLFs at high luminosities is primarily determined by the underlying prior and not by the data itself. However, as we show in Section \ref{sec:diff_nu_flux_XLFs}, the choice of the XLF ultimately has only a minor impact on the resulting diffuse neutrino flux.

For each XLF, we simulate a population of AGNs together with their intrinsic X-ray luminosities. It is particularly important to take into account that the XLF changes not only its normalization but also its shape with redshift. Otherwise, the number of high-luminosity sources at higher redshifts will be severely underestimated. To this end, we consider 14 redshift bins between $z = 0$ and $5$ and fit broken power laws to the XLFs in each bin. We then use the population synthesis framework \texttt{popsynth} \cite{Burgess2021} to simulate three source populations based on the three different XLFs. For each simulated source, the redshift and the intrinsic X-ray luminosity are drawn from the respective distributions.

To ensure a correct description of the number density of bright Seyfert galaxies in the local Universe, we compare our simulated source populations to the BAT AGN Spectroscopic Survey (BASS) catalog \cite{Ricci2017}. This is an X-ray catalog of AGNs detected in the $14 \text{--} 195\,\mathrm{keV}$ band, which has also been used in previous works to predict and search for neutrino emission from nearby Seyfert galaxies \cite{Kheirandish2021, Abbasi2024a, Abbasi2025}. Table \ref{tab:source_pops} lists the number of bright sources with an intrinsic $2 \text{--} 10\,\mathrm{keV}$ X-ray flux of $F_X \geq 10^{-11}\,\mathrm{erg}\,\mathrm{cm}^{-2}\,\mathrm{s}^{-1}$ and the number of sources with an intrinsic X-ray flux as high as or higher than that of NGC 1068 in each simulated population and in the BASS catalog. The U14 XLF and the B15 XLF for the constant-slope prior predict the same order of magnitude of bright, nearby Seyfert galaxies as contained in the BASS catalog. The B15 XLF for the constant-value prior, on the other hand, predicts more than 2 orders of magnitude more sources with $F_X \geq 10^{-11}\,\mathrm{erg}\,\mathrm{cm}^{-2}\,\mathrm{s}^{-1}$, including 46 sources that are as bright as or even brighter than NGC 1068. As mentioned above, this discrepancy can be attributed to the fact that the X-ray sample used to construct the B15 XLFs is not sensitive to rare sources. In line with the findings by Ananna \textit{et al.}~\cite{Ananna2019}, we conclude that the B15 XLF for the constant-value prior likely overestimates the abundance of high-luminosity sources and, therefore, does not provide an accurate description of the number density of bright Seyfert galaxies in the local Universe. 

To correct for this effect and model the contribution of nearby bright sources in a more realistic way, we select all Seyfert galaxies with $F_X \geq 10^{-11}\,\mathrm{erg}\,\mathrm{cm}^{-2}\,\mathrm{s}^{-1}$ from the BASS catalog and add them to our simulated populations. Simultaneously, we remove all simulated sources above this flux threshold to avoid double counting of sources. Overall, we consider 212 Seyfert galaxies from the BASS catalog, including NGC 1068, NGC 4151, and CGCG 420-015. The total number of sources as well as the total intrinsic X-ray flux from all sources in the resulting populations are listed in Table~\ref{tab:source_pops}. The population based on the B15 XLF for the constant-slope prior comprises the largest number of sources, while the population based on the B15 XLF for the constant-value prior produces the highest intrinsic X-ray flux.

% -------------------------------------------------------------------------------------------------

\section{Diffuse neutrino flux}
\label{sec:results}

To estimate the contribution of Seyfert galaxies to the diffuse astrophysical neutrino flux, we apply our neutrino emission model to each individual source in the simulated populations, assuming that the model parameters $P_\mathrm{CR}/P_\mathrm{th}$ and $\eta$ are the same for all sources. First, we fix these two parameters to the best-fit values found for NGC 1068 and study the effect of the different luminosity functions on the resulting diffuse neutrino flux. We then explore how the diffuse flux spectrum depends on the model parameters and quantify the agreement between our diffuse flux predictions and the latest IceCube observations as a function of $P_\mathrm{CR}/P_\mathrm{th}$ and $\eta$. This allows us to derive constraints on these parameters and to assess the exceptional nature of NGC 1068 in the context of the entire population of Seyfert galaxies.

\subsection{Diffuse neutrino flux for different XLFs}
\label{sec:diff_nu_flux_XLFs}

To compute the diffuse neutrino flux from a population of AGNs, we apply the neutrino emission model presented in Section \ref{sec:nu_model} to each source in the population and calculate the neutrino flux taking into account energy losses due to the adiabatic expansion of the Universe. The neutrino flux at Earth from a source located at redshift $z$ is given by
\begin{equation}
\frac{\mathrm{d}N_\nu}{\mathrm{d}E_\nu \mathrm{d}A \mathrm{d}t} (E_\nu) = \frac{(1+z)^2}{4\pi d_L^2 (z)} \frac{\mathrm{d}N_\nu}{\mathrm{d}E_\nu \mathrm{d}t} [E_\nu (1+z)] \,,
\end{equation}
where $\mathrm{d}N_\nu/\mathrm{d}E_\nu \mathrm{d}t$ is the differential neutrino emission at the location of the source given by our neutrino emission model and $d_L (z)$ is the luminosity distance calculated using the approximation from Ref.~\cite{Adachi2012}. To obtain the diffuse neutrino flux per solid angle from the entire population, the neutrino flux spectra of all sources are summed up and divided by $4\pi$ assuming that the sources are isotropically distributed across the sky,
\begin{equation}
\Phi_\nu (E_\nu) = \frac{1}{4\pi} \sum_{i=1}^N \left( \frac{\mathrm{d}N_\nu}{\mathrm{d}E_\nu\mathrm{d}A\mathrm{d}t} \right)_i (E_\nu) \,.
\label{eq:tot_diff_flux}
\end{equation}
Here, $i$ runs over all sources in the population and $(\mathrm{d}N_\nu/\mathrm{d}E_\nu\mathrm{d}A\mathrm{d}t)_i$ is the neutrino flux from the $i$th source.

\begin{figure}[t]
\includegraphics[width=8.5cm]{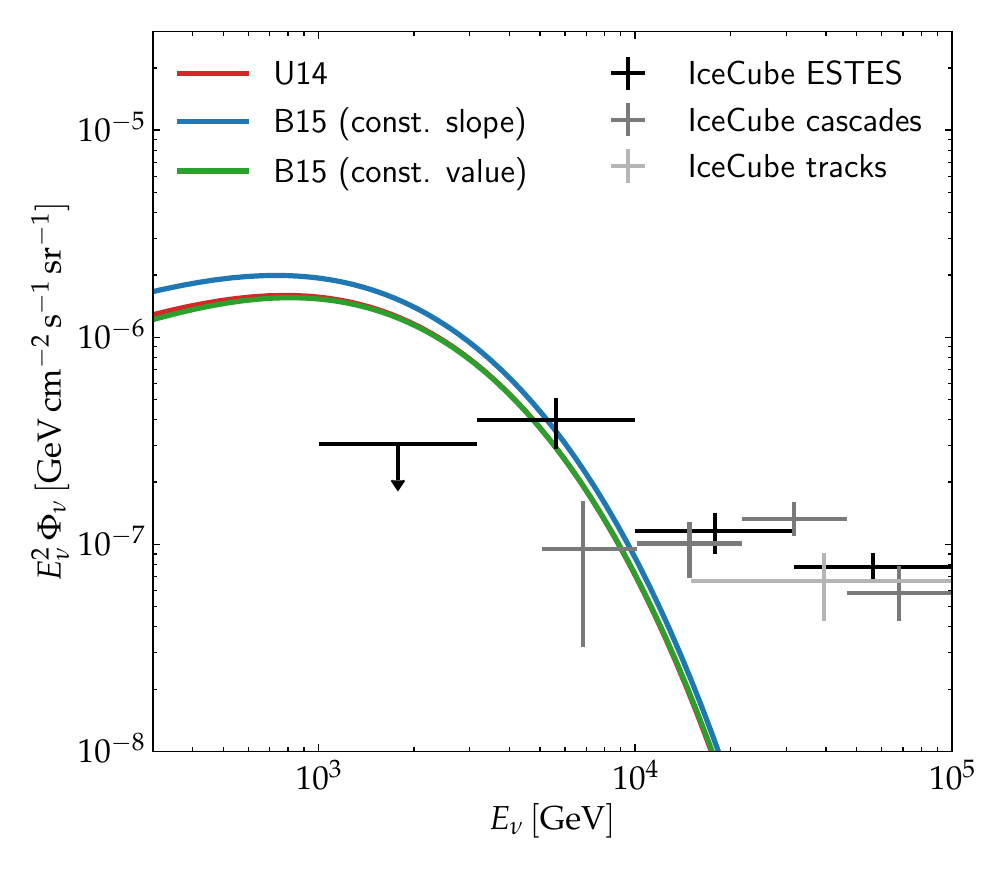}
\caption{Diffuse all-flavor neutrino flux calculated for the three different XLFs assuming $P_\mathrm{CR}/P_\mathrm{th} = 0.5$ and $\eta = 56$ for all sources. The black, dark gray and light gray points correspond to the diffuse neutrino flux measured by IceCube in different analyses assuming an $E^{-2}$ power law in each energy bin \cite{Abbasi2024c, Aartsen2020, Abbasi2022b}. We assume a flavor ratio of $\nu_e : \nu_\mu : \nu_\tau = 1 : 1 : 1$ to convert the per-flavor fluxes reported by IceCube to all-flavor fluxes.}
\label{fig:diff_flux_XLFs}
\end{figure}

\begin{figure*}[t]
\includegraphics[width=17.5cm]{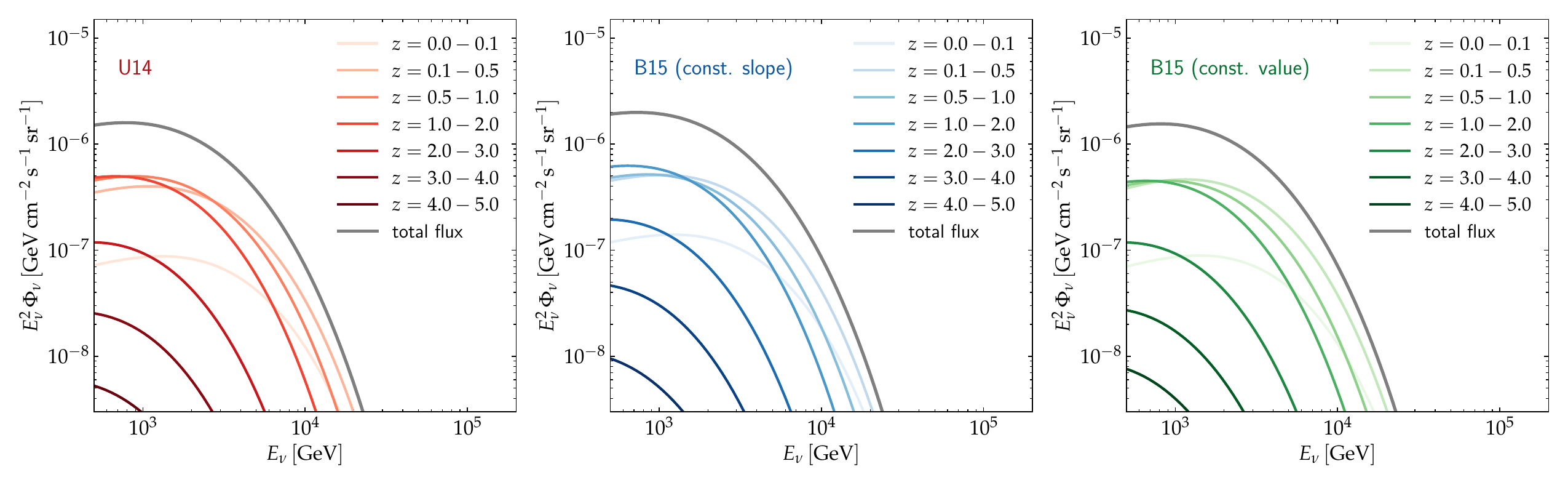}
\caption{Diffuse all-flavor neutrino flux from sources in different redshift bins between $z=0$ and $5$ for the U14 XLF (left panel), the B15 XLF for the constant-slope prior (middle panel) and the B15 XLF for the constant-value prior (right panel) assuming $P_\mathrm{CR}/P_\mathrm{th} = 0.5$ and $\eta = 56$ for all sources. Lighter colors correspond to lower redshifts and the total diffuse flux from all redshift bins together is shown as a gray line.}
\label{fig:diff_flux_z_bins}
\end{figure*}

First, we compute the diffuse neutrino flux assuming that all sources have the same values of the model parameters $P_\mathrm{CR}/P_\mathrm{th}$ and $\eta$ that we found for NGC 1068 in Section \ref{sec:fit} and a coronal radius of $r_c = 5$, since this is the most straightforward way to extrapolate our findings for NGC 1068 to the entire source population. The diffuse flux spectra obtained for the three different XLFs with $P_\mathrm{CR}/P_\mathrm{th} = 0.5$ and $\eta=56$ for all sources are shown in Fig.~\ref{fig:diff_flux_XLFs}. The diffuse neutrino flux measured by IceCube in different analyses using starting tracks and the Enhanced Starting Track Event Selection (ESTES) \cite{Abbasi2024c}, cascade events \cite{Aartsen2020}, and muon tracks from the Northern Hemisphere \cite{Abbasi2022b} are displayed for comparison.

\begin{figure*}[t]
\includegraphics[width=17.5cm]{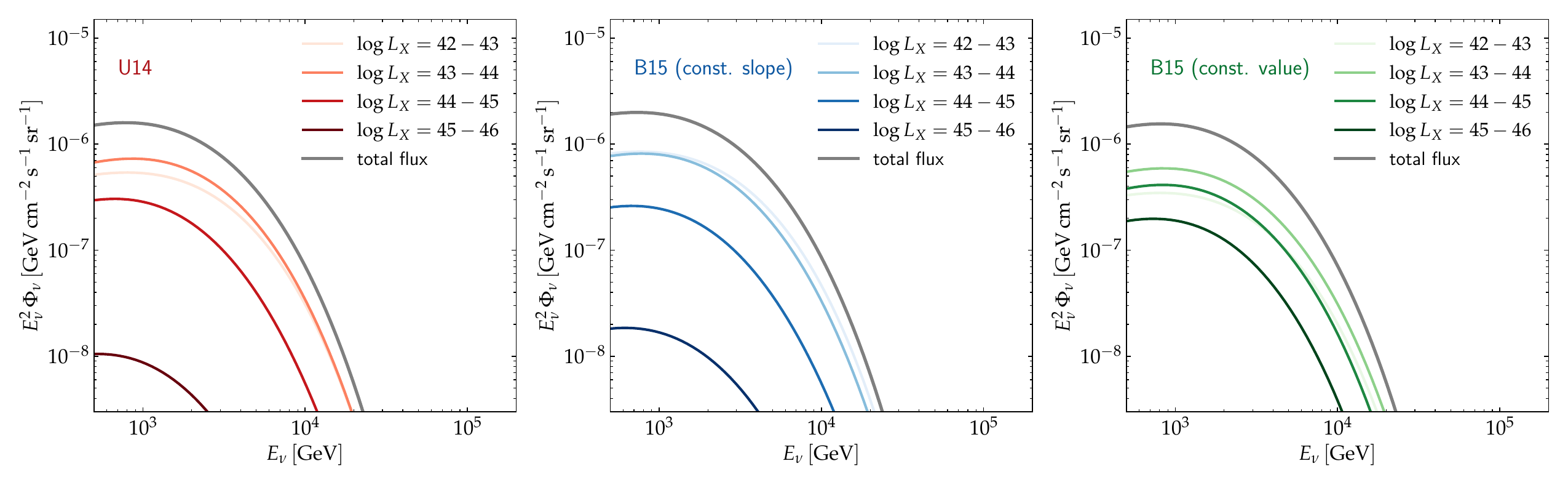}
\caption{Diffuse all-flavor neutrino flux from sources in different X-ray luminosity bins between $\log L_X = 42$ and $46$ for the U14 XLF (left panel), the B15 XLF for the constant-slope prior (middle panel), and the B15 XLF for the constant-value prior (right panel) assuming $P_\mathrm{CR}/P_\mathrm{th} = 0.5$ and $\eta = 56$ for all sources. Lighter colors correspond to lower luminosities, and the total diffuse flux from all luminosity bins together is shown as a gray line.}
\label{fig:diff_flux_lum_bins}
\end{figure*}

Although the underlying XLFs shown in Fig.~\ref{fig:all_LFs} appear quite different, the resulting diffuse neutrino flux spectra are remarkably similar in both shape and overall normalization. They have a slight peak at $\sim\,$1$\,\mathrm{TeV}$ and drop off towards higher energies, with the contribution to the total observed neutrino flux becoming negligible at $E_\nu \gtrsim 10\,\mathrm{TeV}$. The main reason for this similarity is that, after replacing all sources with $F_X \geq 10^{-11}\,\mathrm{erg}\,\mathrm{cm}^{-2}\,\mathrm{s}^{-1}$ by sources from the BASS catalog, the total intrinsic X-ray fluxes of the three populations listed in Table \ref{tab:source_pops} differ by less than a factor of 2, and the neutrino luminosity increases with the X-ray luminosity of a source as described by Eq.~(\ref{eq:Lnu_Lx_scaling}). The source population based on the B15 XLF for the constant-slope prior produces a slightly higher diffuse neutrino flux than the other two populations. This is due to the fact that this population contains the largest number of low-luminosity sources with $L_X \lesssim 10^{43}\,\mathrm{erg}/\mathrm{s}$ and, under the assumption that $P_\mathrm{CR}/P_\mathrm{th}$ is constant across sources of all luminosities, low-luminosity sources have a higher neutrino luminosity relative to their X-ray luminosity than high-luminosity sources. This means that the diffuse neutrino flux is determined not only by the total X-ray flux of the population, but also by the distribution of the source luminosities.

From Fig.~\ref{fig:diff_flux_XLFs} it can also be seen that our diffuse flux predictions for all three XLFs exceed both the 68\% upper limit at $\sim\,$2$\,\mathrm{TeV}$ from the IceCube ESTES analysis and the flux in the lowest energy bin of the IceCube cascade analysis at $\sim\,$7$\,\mathrm{TeV}$. This indicates that it is very unlikely that all sources share the same model parameters as NGC 1068. Especially the assumption that the pressure ratio inside the coronae of all sources reaches the maximum physically allowed value of $P_\mathrm{CR}/P_\mathrm{th} = 0.5$ (see discussion in Section \ref{sec:acceleration}) is quite extreme. Thus, the spectra shown in Fig.~\ref{fig:diff_flux_XLFs} are rather to be understood as upper limits on the possible neutrino emission from the population under the assumption that all sources share the same value of $\eta$ as NGC 1068. A more detailed comparison of our diffuse flux predictions for different values of the model parameters $P_\mathrm{CR}/P_\mathrm{th}$ and $\eta$ to the IceCube observations is performed in Section \ref{sec:IceCube_comparison_diff_flux}.

Another interesting aspect that we investigate is the extent to which sources in different redshift and luminosity bins contribute to the total diffuse neutrino flux from the population. Fig.~\ref{fig:diff_flux_z_bins} shows the relative contribution from sources in different redshift bins to the diffuse flux. For all three XLFs, the bulk of the diffuse flux originates from sources at redshifts $z \,{\sim}\, 0.1\,\text{--}\,2.0$, with sources at higher redshifts contributing to the diffuse flux at lower energies since their neutrino spectra are shifted to lower energies due to the adiabatic expansion of the Universe. At $1\,\mathrm{TeV}$, nearby sources with redshifts $z < 0.1$ account for only $\sim\,$5$\%$ of the diffuse flux due to their limited number. The contribution from the BASS catalog sources, by which we replace all simulated sources with $F_X \geq 10^{-11}\,\mathrm{erg}\,\mathrm{cm}^{-2}\,\mathrm{s}^{-1}$, is even smaller with just $\sim\,$1$\%$ at $1\,\mathrm{TeV}$. Due to their large distances and the fact that the X-ray luminosity density of the AGN population peaks at $z\,{\sim}\,2$ and decreases towards higher redshifts, sources with $z \gtrsim 2$ do not contribute significantly to the diffuse neutrino flux either. 

Fig.~\ref{fig:diff_flux_lum_bins} shows the relative contribution from sources in different X-ray luminosity bins to the diffuse neutrino flux. Considering the populations simulated based on the U14 XLF and the B15 XLF for the constant-slope prior, most of the diffuse flux can be attributed to sources with luminosities in the range $10^{42}\,\mathrm{erg}/\mathrm{s} \lesssim L_X \lesssim 10^{44}\,\mathrm{erg}/\mathrm{s}$. Looking at the population based on the B15 XLF for the constant-value prior, also brighter sources with luminosities up to $L_X \,{\sim}\, 10^{45}\,\mathrm{erg}/\mathrm{s}$ provide a significant contribution because they are so abundant in the population. Under the assumption that all sources are described by the same model parameters as NGC 1068, we conclude that the bulk of the diffuse neutrino flux originates from a large population of very faint neutrino sources and the contribution from nearby Seyfert galaxies is subdominant.

\subsection{Comparison to IceCube observations}
\label{sec:IceCube_comparison_diff_flux}

To constrain our neutrino emission model also from the population side, we quantify the agreement between the diffuse flux predictions and the IceCube measurements as a function of the model parameters $P_\mathrm{CR}/P_\mathrm{th}$ and $\eta$, still assuming that all sources in the population share the same parameter values. The dependence of the shape and the normalization of the diffuse neutrino spectrum on $P_\mathrm{CR}/P_\mathrm{th}$ and $\eta$ is very similar to that of the single-source neutrino spectra shown in Fig.~\ref{fig:neutrino_spectra}. The parameter $\eta$ determines the shape of the spectrum, in particular the cutoff energy, with larger values of $\eta$ leading to lower cutoff energies, while the pressure ratio determines the normalization. We compare our results to the segmented $E^{-2}$ power-law spectra obtained by IceCube in the analyses using ESTES events \cite{Abbasi2024c} and cascade events \cite{Aartsen2020}. The sensitive energy ranges of both analyses extend down to TeV energies, making them well suited to constrain our model. We perform the comparisons to the two IceCube spectra separately, as they differ from each other at energies below $\sim\,$10$\,\mathrm{TeV}$ (see also Fig.~\ref{fig:diff_flux_XLFs}). For the following part of our analysis, we only consider the diffuse flux spectra calculated for the source population based on the U14 XLF, since the results for the three different XLFs do not differ significantly anyway.

We compute the diffuse neutrino flux for a total of 100 values of $\eta$ equally spaced between 1 and 100. In doing so, we only consider sources with $L_X \leq 10^{45.5}\,\mathrm{erg}/\mathrm{s}$ after confirming that the contribution from sources with higher X-ray luminosities is negligible (${<}\,0.25\,\%$) over the entire energy range\footnote{To verify this, we have calculated the diffuse neutrino flux from the full population, including sources with $L_X > 10^{45.5}\,\mathrm{erg}/\mathrm{s}$, on a sparser grid of $\eta$ values ($\eta = 1, 20, 40, 60, 80, 100$), and compared the result to the flux obtained when only considering sources with $L_X \leq 10^{45.5}\,\mathrm{erg}/\mathrm{s}$. Changing the pressure ratio does not affect the relative contribution of sources with different X-ray luminosities, but only the overall normalization of the diffuse flux spectrum, since we assume a common value of $P_\mathrm{CR}/P_\mathrm{th}$ for all sources in the population.}. The reason for this luminosity cut is that calculating the neutrino spectra of sources with high X-ray luminosities is very computationally expensive due to the large size of the corona and the time resolution required for the simulation to capture all relevant interaction processes. 

\begin{figure}[t]
\includegraphics[width=8.5cm]{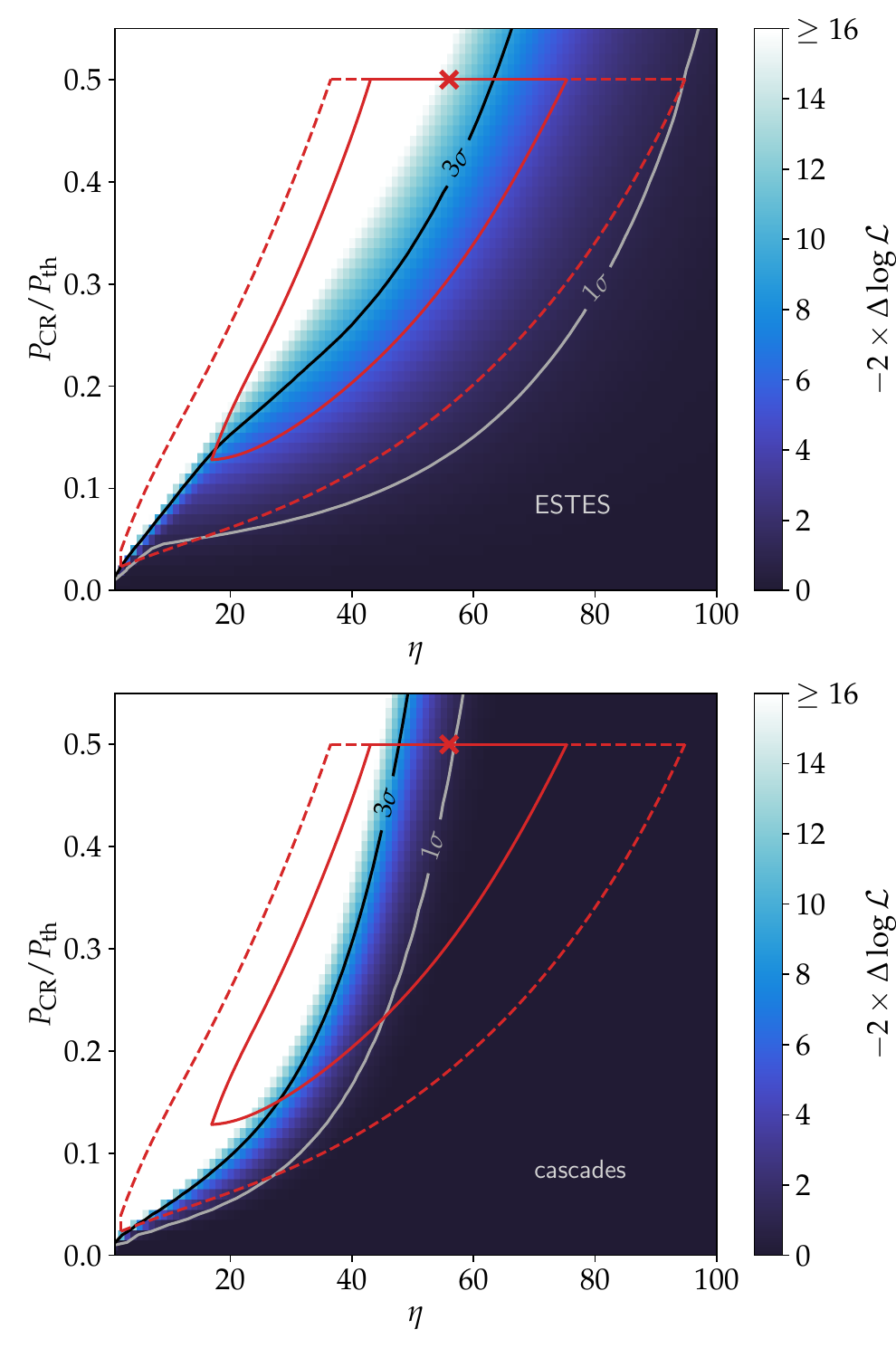}
\caption{Log-likelihood ratio as a function of $P_\mathrm{CR}/P_\mathrm{th}$ and $\eta$ for the comparison of our model predictions for the diffuse neutrino flux to the IceCube measurements using ESTES (top panel) and cascade (bottom panel) events. The gray and black contours indicate the $1\sigma$ and $3\sigma$ exclusion regions. The red solid and dashed contours represent the $1\sigma$ and $2\sigma$ confidence regions from the fit of our model neutrino spectrum for NGC 1068 to the public IceCube data, and the red cross corresponds to the best-fit result of $P_\mathrm{CR}/P_\mathrm{th} = 0.5$ and $\eta = 56$.}
\label{fig:contour_plots_diff_flux}
\end{figure}

In order to compare our diffuse flux spectra to the segmented $E^{-2}$ power-law fluxes reported by IceCube, we calculate the average all-flavor neutrino flux in the $i$th energy bin as
\begin{equation}
\Phi_i = \frac{1}{E_{i+1} - E_i} \int_{E_i}^{E_{i+1}}\mathrm{d}E_\nu\, E_\nu^2\, \Phi_\nu (E_\nu)\,,
\end{equation}
where $\Phi_\nu (E_\nu)$ is the neutrino flux from the population given by Eq.~(\ref{eq:tot_diff_flux}), and $E_i$ and $E_{i+1}$ are the lower and the upper bound of the energy bin, respectively. For each bin, we define a likelihood of the form 
\begin{equation}
\mathcal{L} (\Phi_i | \Phi_{\mathrm{obs}, i}, \sigma_i) \propto
\begin{cases}
\frac{1}{\sigma_i \sqrt{2\pi}} & \mathrm{if}\,\, \Phi_i < \Phi_{\mathrm{obs}, i} \\
\mathcal{N} (\Phi_i | \Phi_{\mathrm{obs}, i}, \sigma_i) & \mathrm{if}\,\, \Phi_i \geq \Phi_{\mathrm{obs}, i}
\end{cases} 
\end{equation}
that is constant for $\Phi_i < \Phi_{\mathrm{obs}, i}$ and has the shape of a Gaussian with
\begin{equation}
\mathcal{N}(x|\mu, \sigma) \equiv \frac{1}{\sigma\sqrt{2\pi}} e^{-\frac{(x-\mu)^2}{2\sigma^2}}
\end{equation}
for $\Phi_i \geq \Phi_{\mathrm{obs}, i}$. The flux and the corresponding 68\% uncertainty in the $i$th energy bin reported by IceCube are denoted by $\Phi_{\mathrm{obs}, i}$ and $\sigma_i$, respectively, where we assume a flavor ratio of $\nu_e : \nu_\mu : \nu_\tau = 1 : 1 : 1$ to convert per-flavor fluxes to all-flavor fluxes. The likelihood for a total of $N$ energy bins is then given by
\begin{equation}
\mathcal{L}(\Phi|\Phi_\mathrm{obs}, \sigma) = \prod_{i=1}^N \mathcal{L}(\Phi_i|\Phi_{\mathrm{obs}, i}, \sigma_i)
\end{equation}
with $\Phi \equiv (\Phi_1, \dots, \Phi_N)$, $\Phi_\mathrm{obs} \equiv (\Phi_{\mathrm{obs}, 1}, \dots, \Phi_{\mathrm{obs}, N})$ and $\sigma \equiv (\sigma_1, \dots, \sigma_N)$. Upper limits are treated as flux measurements with $\Phi_{\mathrm{obs}, i} = 0$ and $\sigma_i$ as reported in the respective publications.

Fig.~\ref{fig:contour_plots_diff_flux} shows the log-likelihood ratio
\begin{equation}
-2\times \Delta\log\mathcal{L} \equiv -2\log \frac{\mathcal{L}(\Phi|\Phi_\mathrm{obs}, \sigma)}{\mathcal{L}(\Phi = 0|\Phi_\mathrm{obs}, \sigma)} \,,
\end{equation}
where $\Phi \equiv \Phi (P_\mathrm{CR}/P_\mathrm{th}, \eta)$ is a function of the model parameters $P_\mathrm{CR}/P_\mathrm{th}$ and $\eta$. The regions of the parameter space that lie above the gray and black contours are excluded by the IceCube measurements with a significance of $1\sigma$ and $3\sigma$, respectively. These exclusion contours correspond to  $-2\times\Delta\log\mathcal{L} = 1$ and $9$ according to Wilks' theorem \cite{Wilks1938}, the use of which is justified since we assume that the uncertainties on the segmented power law fluxes reported by IceCube are Gaussian. The red solid and dashed contours correspond to the $1\sigma$ and $2\sigma$ contours shown in Fig.~\ref{fig:contour_plot_NGC1068} with the signal event number $n_s$ converted to $P_\mathrm{CR}/P_\mathrm{th}$. For the visualization in Fig.~\ref{fig:contour_plots_diff_flux}, we use a polynomial fit to smooth the contours that are calculated on a finite grid of values for $n_s$ and $\eta$. The sharp cutoff at $P_\mathrm{CR}/P_\mathrm{th} = 0.5$ is due to the fact that, for higher pressure ratios, the AGN corona is no longer stable, as explained in Section \ref{sec:acceleration}. 

\begin{figure}[t]
\includegraphics[width=8.5cm]{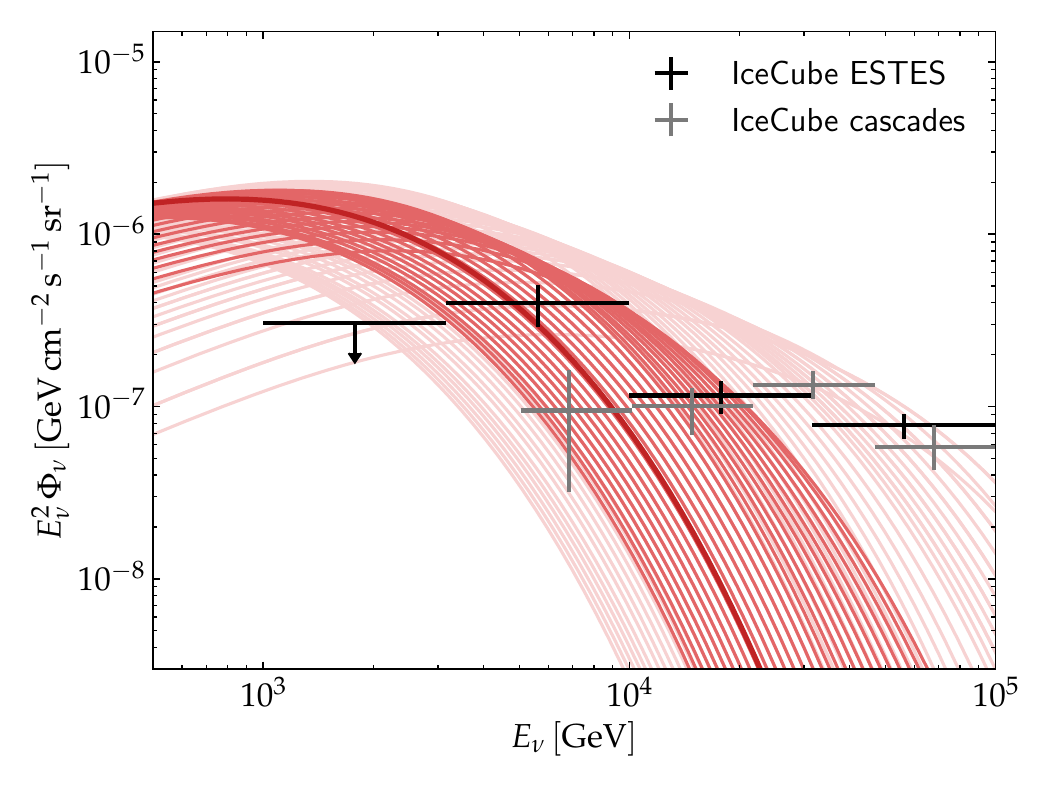}
\caption{Diffuse all-flavor neutrino flux for $P_\mathrm{CR}/P_\mathrm{th} = 0.5$ and $\eta = 56$ (thick dark red line) together with discrete shaded spectra indicating the $1\sigma$ and $2\sigma$ confidence regions of the fit for NGC 1068. Shown is a selection of spectra for parameter values that lie within the red solid and dashed contours plotted in Fig.~\ref{fig:contour_plots_diff_flux}. The black and dark gray points correspond to the diffuse flux measured by IceCube using starting-track \cite{Abbasi2024c} and cascade \cite{Aartsen2020} events, respectively.}
\label{fig:diff_flux_error_band}
\end{figure}

Comparing our diffuse flux predictions based on the neutrino emission model presented in Section \ref{sec:nu_model} to the IceCube observations, we can exclude scenarios in which all sources in the population have a high pressure ratio in combination with a high level of turbulence inside the corona, i.e.~a high value of $P_\mathrm{CR}/P_\mathrm{th}$ together with a small value of $\eta$. Due to the upper limit at $\sim\,$2$\,\mathrm{TeV}$ and the shape of the diffuse neutrino spectra predicted by our model, the IceCube ESTES measurement is more constraining than the one based on cascade events. As can be seen in the upper panel in Fig.~\ref{fig:contour_plots_diff_flux}, the best-fit result for NGC 1068 indicated by a red cross lies outside the $3\sigma$ contour. If all sources in the population had the same model parameters as NGC 1068, the resulting diffuse neutrino flux would exceed the diffuse flux from the ESTES analysis by $3.8\sigma$, echoing our findings in Fig.~\ref{fig:diff_flux_XLFs}. Looking at the contours that indicate the $1\sigma$ and $2\sigma$ confidence regions of the fit for NGC 1068, it becomes apparent that it is challenging to find values of $P_\mathrm{CR}/P_\mathrm{th}$ and $\eta$ for which the resulting diffuse flux spectrum agrees with the IceCube ESTES flux while the single-source neutrino spectrum for NGC 1068 still provides a good fit to the public IceCube data. This is also illustrated in Fig.~\ref{fig:diff_flux_error_band}, which shows a selection of diffuse neutrino spectra for parameter values within the $1\sigma$ and $2\sigma$ confidence regions of the fit result for NGC 1068 marked in Fig.~\ref{fig:contour_plots_diff_flux}. Here, the upper limit at $\sim\,$2$\,\mathrm{TeV}$ lies just at the edge of the $2\sigma$ band.

As shown in the bottom panel of Fig.~\ref{fig:contour_plots_diff_flux}, the constraints imposed by the diffuse flux spectrum derived from cascade events are weaker, mainly because its sensitive energy range starts only at $\sim\,$7$\,\mathrm{TeV}$. Our model spectrum for the best-fit parameters obtained for NGC 1068 exceeds the observed flux by only $1.4\sigma$. In addition, there is a range of  parameter values for which our model predictions for both the single-source spectrum of NGC 1068 and the diffuse flux are in agreement with the observations. This can also be seen in Fig.~\ref{fig:diff_flux_error_band}, where the segmented power-law flux derived from cascade events lies within or above the $1\sigma$ band in all energy bins.

Here, we only compare our results to the IceCube measurements using starting-track and cascade events and do not perform a similar comparison to the one using through-going muon tracks \cite{Abbasi2022b}. The sensitive energy range of this analysis only starts at $15\,\mathrm{TeV}$, but our diffuse neutrino spectra peak at energies ${\lesssim}\,10\,\mathrm{TeV}$. Therefore, the diffuse flux spectrum derived from through-going muon tracks is not suited to constrain our model. At energies below $\sim\,$100$\,\mathrm{TeV}$, the result of the combined fit using cascades and muon tracks presented in Ref.~\cite{Naab2023} is very similar to the spectrum obtained in the cascades-only analysis, since both analyses use the same underlying cascade event sample and the cascade events drive the combined fit at low energies. Thus, comparing our diffuse flux predictions to the combined fit spectrum would yield a result almost identical to that of the comparison to the spectrum derived from cascade events only.

% -------------------------------------------------------------------------------------------------

\section{Discussion}
\label{sec:discussion}

Our results show that the neutrino observations of NGC 1068 are well described by our neutrino emission model, which is based on the disk-corona model presented in Ref.~\cite{Murase2020}. However, extrapolating our model to the entire population of Seyfert galaxies, assuming that all sources share the same parameters as NGC 1068, yields a diffuse neutrino flux that exceeds current upper limits at TeV energies. This suggests that only a small fraction of Seyfert galaxies, at most $\sim\,$25$\%$\footnote{Under the assumption that the remaining sources do not emit any neutrinos, up to $\sim\,$25$\%$ of all sources could have the same model parameters as NGC 1068 without the resulting diffuse neutrino flux exceeding the IceCube observations by more than $1\sigma$.}, can be as efficient neutrino emitters as NGC 1068.

\subsection{Coronal properties of NGC 1068}

As shown in Section \ref{sec:fit_nu_spectrum}, the best-fit model parameters obtained for NGC 1068 lie at the upper edge of the physically allowed parameter space. For the normalization of the neutrino spectrum, our fit yields the maximum allowed value for the pressure ratio of $P_\mathrm{CR}/P_\mathrm{th} = 0.5$, indicating that the corona of NGC 1068 must host an extremely efficient hadronic particle accelerator for our model to explain the observed neutrino emission (see also discussion in Section \ref{sec:acceleration}). This is in agreement with the results of previous studies, which compare the predictions of different neutrino emission models to the IceCube power-law spectrum of NGC 1068 and find similarly high pressure ratios \cite{Murase2020, Kheirandish2021, Eichmann2022, Blanco2023, Das2024}. Abbasi \textit{et al.}~\cite{Abbasi2024a} fit the model for NGC 1068 and also other nearby Seyfert galaxies presented in Ref.~\cite{Kheirandish2021} to the proprietary IceCube data consisting of 11 years of through-going muon tracks from the Northern sky. Fitting only the normalization and keeping the overall shape of the neutrino spectrum fixed, they also find a pressure ratio of $P_\mathrm{CR}/P_\mathrm{th} \simeq 0.5$ for NGC 1068.
 
The size of the corona is strongly constrained by the Fermi-LAT observations of NGC 1068 \cite{Ajello2023}. As shown in Section \ref{sec:gamma_ray_spectrum_NGC1068}, in order for our neutrino emission model for NGC 1068 not to exceed the upper limit on the $\gamma$-ray flux at $\sim\,$30$\,\mathrm{MeV}$, we need a coronal radius of $R_c \lesssim 5R_S$. This is consistent with the findings of Das \textit{et al.} \cite{Das2024}, who obtain $R_c \lesssim (3-15) R_S$ depending on the magnetic field strength inside the corona. In general, such small and compact coronae are consistent with the results of various observational methods used to estimate the size of AGN coronae \cite{Laha2025}. Thus, not only neutrino data but also $\gamma$-ray observations in the MeV range are essential for a better understanding of AGN coronae as possible neutrino production sites.

\subsection{Description of the number density of X-ray bright AGNs}

As described in Section \ref{sec:population}, the predictions of the U14 XLF and the B15 XLF for the constant-slope prior agree with the BASS catalog within one order of magnitude (see also Table \ref{tab:source_pops}). Consequently, replacing the simulated bright sources with sources from the BASS catalog has a negligible effect on the resulting diffuse neutrino flux. This is not the case for the B15 XLF for the constant-value prior, which predicts more than two orders of magnitude more sources with $F_X \geq 10^{-11}\,\mathrm{erg}\,\mathrm{cm}^{-2}\,\mathrm{s}^{-1}$ than are present in the BASS catalog. In particular, the simulated population contains 46 sources that are as bright or brighter than NGC 1068, some even with intrinsic X-ray fluxes of $F_X \geq 10^{-8}\,\mathrm{erg}\,\mathrm{cm}^{-2}\,\mathrm{s}^{-1}$. Assuming that these sources are described by similar model parameters as NGC 1068, at least some of them should have already been discovered by IceCube. Since this is not the case, we conclude that of the two XLFs derived by Buchner \textit{et al.}, the one for the constant-slope prior provides a more realistic description of the number density of AGNs, especially at low redshifts and for high X-ray luminosities. This is consistent with the findings of Refs.~\cite{Buchner2015} and \cite{Ananna2019}.

\subsection{Implications of our diffuse flux predictions}
 
Our results for all three XLFs indicate that Seyfert galaxies can account for the diffuse astrophysical neutrino flux at energies below $\sim\,$10$\,\mathrm{TeV}$ without exceeding their energy budget. To explain the observed diffuse neutrino flux at higher energies, at least one additional population of neutrino sources, such as starburst galaxies \cite{Peretti2020, Ambrosone2021} or AGN outflows \cite{Ehlert2025, Abbasi2022c}, is needed. A study of different populations of candidate neutrino sources presented in Ref.~\cite{Groth2025} comes to the conclusion that low-luminosity AGNs, misaligned radio-loud AGNs, and galaxy clusters could also contribute significantly to the diffuse neutrino flux. The possible contribution from blazars, on the other hand, is limited to ${\lesssim}\,10\%$ at $100\,\mathrm{TeV}$ \cite{Aartsen2017, Huber2019}. For further details on the contribution of blazars to the astrophysical neutrino flux, see \cite{Giommi2021} and references therein.

Evaluating how well our model predictions agree with the diffuse flux measured by IceCube, we obtain detailed constraints on a neutrino emission model for Seyfert galaxies based on both observations of NGC 1068 as a nearby point source candidate and the cumulative neutrino emission of the entire source population. Our results show that if all sources were described by the same model parameters as NGC 1068, their neutrino emission would significantly exceed current diffuse flux measurements at TeV energies, in particular the upper limit at ${\sim}\, 2\,\mathrm{TeV}$ from the IceCube ESTES analysis \cite{Abbasi2024c}. 

\begin{figure*}[t]
\includegraphics[width=17.5cm]{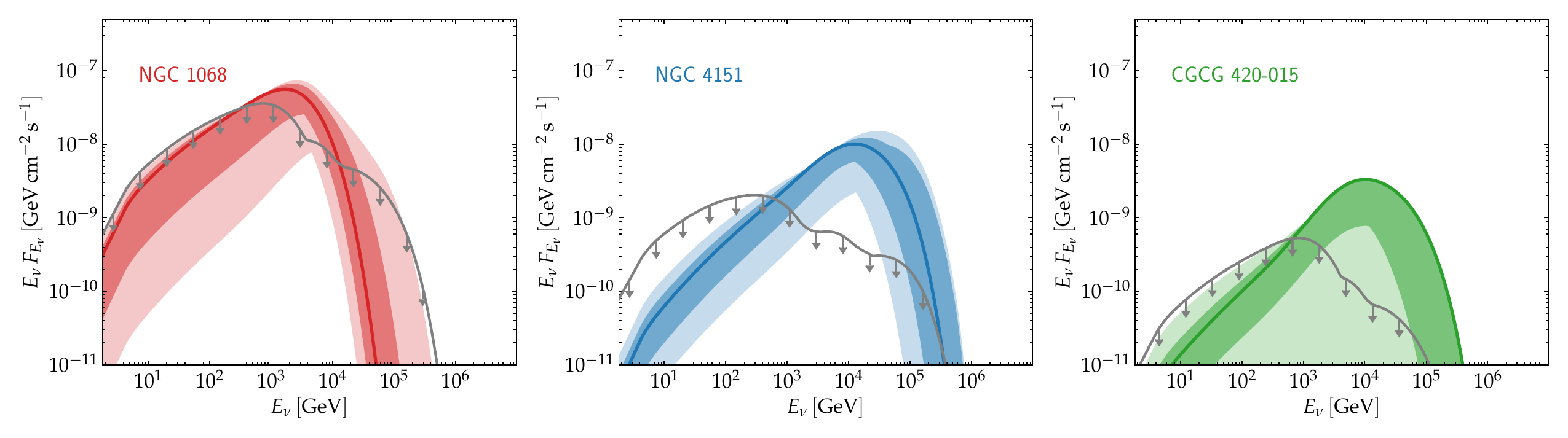}
\caption{Best-fit all-flavor neutrino spectra of NGC 1068 (left panel), NGC 4151 (middle panel), and CGCG 420-015 (right panel) obtained using \texttt{SkyLLH}. The shaded bands correspond to the $1\sigma$ and $2\sigma$ uncertainty regions. For CGCG 420-015, the $2\sigma$ band extends down to a neutrino flux of zero, reflecting that the neutrino excess associated with this source is not statistically significant and a scenario with no associated signal events ($n_s = 0$) is consistent with the IceCube data within $2\sigma$. The corresponding best-fit values of the model parameters are listed in Table \ref{tab:fit_results_nearby_sources}. For each source, the gray line indicates the maximum allowed neutrino emission for which the corresponding diffuse flux would still be compatible with the IceCube observations within $1\sigma$.}
\label{fig:best_fit_spectra_nearby_sources}
\end{figure*}

\begin{table*}[t]
\caption{Distances and intrinsic X-ray luminosities as well as best-fit model parameters for the nearby Seyfert galaxies NGC 1068, NGC 4151, and CGCG 420-015. The total CR proton luminosity needed to achieve the given pressure ratio is denoted by $L_p$.}
\begin{ruledtabular}
\begin{tabular}{lclccccc}
& $d_L\,[\mathrm{Mpc}]$ & $L_X\, [\mathrm{erg}/\mathrm{s}]$ & $\hat{n}_s$ & $\hat{\eta}$ & $P_\mathrm{CR}/P_\mathrm{th}$ & $L_p\, [\mathrm{erg}/\mathrm{s}]$ & $L_p/L_\mathrm{Edd}$ \\
\hline
NGC 1068 & 11.14 & $4.2\times 10^{43}$ & 46.4 & 56 & 0.5 & $6.0\times 10^{43}$ & 0.009 \\
NGC 4151 & 15.8\footnotemark[1] & $2.5\times 10^{42}$\footnotemark[2] & 23.5 & 4 & 0.5 & $1.7\times 10^{43}$ & 0.020 \\
CGCG 420-015 & 128.8\footnotemark[3] & $1.0\times 10^{44}$\footnotemark[3] & 8.6 & 1 & 0.5 & $4.7\times 10^{44}$ & 0.038 \\
\end{tabular}
\end{ruledtabular}
\footnotetext[1]{From \cite{Yuan2020}.}
\footnotetext[2]{Calculated from the intrinsic $2\,\text{--}\,10\,\mathrm{keV}$ X-ray flux reported in \cite{Ricci2017}.}
\footnotetext[3]{From \cite{Ricci2017}.}
\label{tab:fit_results_nearby_sources}
\end{table*}

For the population of Seyfert galaxies to account for the bulk of the observed diffuse flux below $\sim\,$100$\,\mathrm{TeV}$ without violating this upper limit, the majority of sources would have to have a much lower pressure ratio of $P_\mathrm{CR}/P_\mathrm{th} \,{\sim}\, 0.01$ and a very high level of turbulence with $\eta \,{\sim}\, 1$. However, as indicated by the red solid and dashed contours in Fig.~\ref{fig:contour_plots_diff_flux}, the corresponding single-source neutrino spectrum for NGC 1068 is not compatible with the IceCube data. It is generally not possible to find values of the model parameters $P_\mathrm{CR}/P_\mathrm{th}$ and $\eta$ for which our model provides a good fit to the public IceCube data for NGC 1068 without significantly overproducing the observed diffuse neutrino flux. Thus, we conclude that NGC 1068 is an extraordinarily powerful Seyfert galaxy and that most Seyfert galaxies in the Universe must be significantly less efficient neutrino emitters.
 
A similar conclusion can be drawn for two other nearby Seyfert galaxies, NGC 4151 and CGCG 420-015, for which the IceCube Collaboration has also reported evidence for neutrino emission \cite{Abbasi2024a, Abbasi2025}. In addition, neutrino and $\gamma$-ray emission from NGC 4151 was studied by~\cite{Murase:2023ccp}. Following the procedure described in Section \ref{sec:fit_nu_spectrum} for NGC 1068, we fit our neutrino emission model for these two sources to the public IceCube data. The best-fit neutrino spectra of NGC 4151 and CGCG 420-015 shown in Fig.~\ref{fig:best_fit_spectra_nearby_sources} peak at slightly higher energies than that of NGC 1068, but have a lower overall normalization. The obtained number of signal events and the corresponding best-fit parameter values are listed in Table \ref{tab:fit_results_nearby_sources}. For both sources, our fit yields the maximum allowed value of the pressure ratio of $P_\mathrm{CR}/P_\mathrm{th} = 0.5$, but smaller values of $\eta$ than for NGC 1068, namely $\eta = 4$ for NGC 4151 and $\eta = 1$ for CGCG 420-015.

However, as shown in Fig.~\ref{fig:contour_plots_diff_flux}, the resulting diffuse neutrino flux for these parameter values would dramatically exceed the IceCube observations. This is further illustrated by the fact that the best-fit neutrino spectra for both NGC 4151 and CGCG 420-015 lie well above the gray lines shown in Fig.~\ref{fig:best_fit_spectra_nearby_sources}, which indicate the maximum neutrino emission compatible with the observed diffuse flux. Values of the model parameters $P_\mathrm{CR}/P_\mathrm{th}$ and $\eta$ for which the single-source neutrino spectra lie fully below the gray line result in diffuse neutrino spectra that exceed the IceCube measurements by less than $1\sigma$. So, if we assumed that all sources were described by the same model parameters as NGC 4151 or CGCG 420-015, the tension between our diffuse flux predictions and the IceCube observations would become even stronger. Padovani \textit{et al.}~\cite{Padovani2024b} come to a similar conclusion based on the fact that NGC 1068 has a much lower neutrino-to-X-ray ratio than both NGC 4151 and CGCG 420-015. 

Thus, those Seyfert galaxies that start to emerge as neutrino point sources in the IceCube data appear to be above-average neutrino emitters, and it is very unlikely that all sources in the population share the same neutrino emission properties. To obtain a more realistic estimate of the contribution of Seyfert galaxies to the diffuse neutrino flux, it is necessary to assume a distribution of $P_\mathrm{CR}/P_\mathrm{th}$ and $\eta$ among the population. For Seyfert galaxies to be able to account for the observed diffuse flux up to ${\sim}\,100\,\mathrm{TeV}$ without exceeding the upper limit at ${\sim}\,2\,\mathrm{TeV}$, a distribution for which the majority of sources have a low pressure ratio ($P_\mathrm{CR}/P_\mathrm{th} \lesssim 0.05$) and a high level of turbulence ($\eta \lesssim 10$) is needed. The resulting diffuse neutrino spectra for two distributions that fulfill this requirement, as well as for a uniform distribution of $P_\mathrm{CR}/P_\mathrm{th}$ and $\eta$, are shown in Appendix \ref{sec:diff_flux_distributions}. Nevertheless, the exact form of such a distribution remains unknown.

In a stacking analysis, the IceCube Collaboration has derived an upper limit on the diffuse neutrino emission from Seyfert galaxies, which provides slightly stronger constraints than the total observed diffuse flux at energies higher than $\sim\,$20$\, \mathrm{TeV}$ \cite{Abbasi2025}. However, this limit was calculated under the assumption that the neutrino spectra of all sources follow a single power law with a fixed spectral index and that the neutrino luminosity of a source is proportional to its X-ray luminosity. Therefore, a direct comparison with predictions of other models is only possible to a limited extent.

Murase \textit{et al.}~\cite{Murase2020}, whose description of AGN coronae our neutrino emission model is based on, have also studied the contribution of Seyfert galaxies to the diffuse neutrino flux. Their model neutrino spectrum for NGC 1068 has a lower normalization and a higher cutoff energy than ours, reflecting their focus on explaining the high-energy end of the observed neutrino emission. Consequently, their diffuse neutrino spectrum also has a lower normalization and peaks at higher energies than our result for the best-fit parameters for NGC 1068, shown in Fig.~\ref{fig:diff_flux_XLFs}. This allows them to reproduce the observed diffuse flux in the ${\sim}\, 7\, \text{--}\, 70\,\mathrm{TeV}$ range while remaining consistent with observations of the diffuse MeV $\gamma$-ray background. Another difference lies in the treatment of the non-thermal proton spectrum. We approximate it as a power law with an exponential cutoff (see Section \ref{sec:acceleration}), whereas Murase \textit{et al.}~solve the Fokker-Planck equation. For sources with low $L_X$ and high $\eta$, for which proton acceleration is limited by diffusive escape, this yields cutoff energies that are ${\sim}\,10\,\text{--}\,30$ times higher than those that we obtain by equating the acceleration time and the total loss time \cite{Murase:2023ccp}. Consequently, this leads to a slightly different mapping between $\eta$ and the cutoff energy of the proton and neutrino spectra in our framework. Moreover, for sources with high $L_X$ and low $\eta$, for which acceleration is stopped by $p\gamma$ and Bethe-Heitler interactions, Murase \textit{et al.}~obtain very peaked spectra with a pile-up just below the cutoff. Thus, in our model, these sources require somewhat higher CR pressures to achieve the same peak neutrino flux.  

Padovani \textit{et al.}~\cite{Padovani2024b} estimated the diffuse flux contribution from non-jetted AGNs based on the model neutrino spectrum for NGC 1068 presented in Ref.~\cite{Murase2022}, simply assuming that the neutrino flux of a source scales linearly with its intrinsic X-ray flux. In contrast, under the assumption that the CR-to-thermal pressure ratio is constant among all sources, we find that the neutrino luminosity increases slower than linearly with the X-ray luminosity (see Eq.~(\ref{eq:Lnu_Lx_scaling})). With this assumption, we also ensure that all sources have a pressure ratio lower than $P_\mathrm{CR}/P_\mathrm{th} = 0.5$, above which the AGN corona becomes unstable (see Section \ref{sec:acceleration}).

Using a neutrino emission model based on non-resonant stochastic acceleration in strong turbulence with a much higher level of magnetization than in our model, Fiorillo \textit{et al.}~\cite{Fiorillo2025} can explain the observed diffuse neutrino flux in the ${\sim}\, 10\,\text{--}\,100\,\mathrm{TeV}$ energy range. In addition, they find a scaling of $L_\nu$ with $L_X$ that is similar to ours. However, contrary to our work, they do not quantify the consistency of their model with the IceCube observations of NGC 1068. Our approach is also different from that by Ambrosone \cite{Ambrosone2024}. Fitting a simple analytical leaky-box model to the neutrino luminosities of four nearby Seyfert galaxies, they obtain a diffuse neutrino flux that exceeds the IceCube observations below $\sim\,$30$\,\mathrm{TeV}$ by almost one order of magnitude. 

% -------------------------------------------------------------------------------------------------

\section{Conclusions}
\label{sec:conclusions}

We have investigated the potential contribution of Seyfert galaxies to the astrophysical diffuse neutrino flux, assuming that inside AGN coronae, protons undergo stochastic acceleration to energies of a few hundred TeV and subsequently interact with the surrounding medium producing both neutrinos and $\gamma$-rays. Fitting our neutrino emission model to the public IceCube data, we find that explaining the observed neutrino flux from the nearby Seyfert galaxy NGC 1068 requires this source to be an exceptionally efficient hadronic accelerator characterized by the maximum allowed value of the CR-to-thermal pressure ratio of $P_\mathrm{CR}/P_\mathrm{th} = 0.5$ and an inverse turbulence strength of $\eta = 56$. In addition, we were able to constrain the size of the corona to less than 5 Schwarzschild radii by comparing the predicted cascaded $\gamma$-ray emission to the latest Fermi-LAT observations of NGC 1068. 

To estimate the total diffuse neutrino flux from Seyfert galaxies, we have extrapolated our model to the full source population based on the X-ray luminosity function of AGNs. Comparing our results to observations, we have placed the first detailed constraints on a physically motivated neutrino emission model for Seyfert galaxies informed by neutrino and $\gamma$-ray data of NGC 1068 as a nearby benchmark source as well as measurements of the diffuse neutrino flux, accurately incorporating uncertainties. 

Our results indicate that Seyfert galaxies can account for a sizable fraction of the observed diffuse neutrino flux below $\sim\,$10$\,\mathrm{TeV}$. However, assuming that all sources share the same model parameters as NGC 1068 leads to an overproduction of the observed diffuse flux by $3.8\sigma$. Explaining the neutrino emission of NGC 4151 and CGCG 420-015, two other nearby Seyfert galaxies that are beginning to emerge as neutrino point sources, requires even more extreme parameter values for which the tension between our model predictions and the observed diffuse flux would become even stronger. This suggests that those Seyfert galaxies for which IceCube starts to see evidence for neutrino emission are not representative of the broader population, but rather outliers with unusually efficient neutrino production. Under the assumption of the neutrino emission model presented in Section \ref{sec:nu_model}, we can rule out scenarios in which the majority of sources have a high CR pressure ($P_\mathrm{CR}/P_\mathrm{th} \gtrsim 0.2$) in combination with a high level of turbulence ($\eta \lesssim 40$). 

Neutrino observations of other nearby Seyfert galaxies, including those in the Southern sky, will be particularly important to gain a better understanding of the uniqueness of NGC 1068 in the context of the entire population. Next-generation neutrino telescopes such as IceCube-Gen2 and KM3NeT will play a key role in this. IceCube-Gen2 is a planned expansion of the instrumented volume of IceCube from $1\,\mathrm{km}^3$ to $8\,\mathrm{km}^3$ \cite{Aartsen2021}. This will increase the effective area for through-going muon tracks by a factor of 5 \cite{Clark2021} and improve the angular resolution since longer muon tracks can be contained within the detector volume. KM3NeT is a cubic kilometer neutrino telescope currently under construction in the Mediterranean \cite{AdrianMartinez2016}. Using seawater as a detection medium, it will have a better angular resolution than IceCube due to the longer scattering length of Cherenkov photons. In addition, KM3NeT will be more sensitive to neutrino point sources in the Southern sky due to its location in the Northern Hemisphere. This is particularly interesting since several Seyfert galaxies located in the Southern sky are intrinsically very bright in X-rays. Assuming that the bulk of the observed X-ray flux is produced in the AGN corona, the most promising targets for KM3NeT include the Circinus galaxy, ESO 138-G001, NGC 7582, and Centaurus A \cite{Ricci2017, Kheirandish2021}. Baikal-GVD, which is currently under construction in Lake Baikal \cite{Aynutdinov2023}, and planned neutrino telescopes such as P-ONE in the Northeast Pacific Ocean \cite{Agostini2020}, TRIDENT in the South China Sea \cite{Ye2023}, and HUNT \cite{Huang2023} will further enhance the chances of identifying neutrino point sources in both the Northern and the Southern sky.

In addition, $\gamma$-ray observations at MeV energies provide a means to test the neutrino emission model underlying our calculations. If Seyfert galaxies contribute to the diffuse neutrino flux at TeV energies, they may also account for a significant fraction of the diffuse MeV $\gamma$-ray background \cite{Murase2020, Inoue2019}. Moreover, observations of nearby individual sources in the MeV band can be used to further constrain possible neutrino production sites. Future MeV $\gamma$-ray telescopes such as newASTROGAM \cite{Berge2025}, AMEGO-X \cite{Caputo2022}, and COSI \cite{Tomsick2023} will therefore be crucial for improving our understanding of Seyfert galaxies as hidden neutrino sources.

% -------------------------------------------------------------------------------------------------

\begin{acknowledgments}
The authors would like to thank Martina Karl and Chiara Bellenghi for their support with using  \texttt{SkyLLH} to fit our model to the public IceCube data. We also thank Xavier Rodrigues for insight into incorporating the diffusive escape mechanism (see Section~\ref{sec:losses}) into the \texttt{AM\textsuperscript{3}} framework. Furthermore, we are grateful to Martin Lemoine for valuable discussions on particle acceleration in magnetized turbulence.
\end{acknowledgments}

% -------------------------------------------------------------------------------------------------

\appendix

\section{Proton timescales}
\label{sec:proton_timescales}

As explained in Section \ref{sec:losses}, non-thermal protons inside AGN coronae can lose energy via $pp$ interactions, $p\gamma$ interactions, Bethe-Heitler pair production, and synchrotron radiation. In addition, they can escape from the corona via infall onto the SMBH and diffusion. 

\begin{figure*}
\includegraphics[width=17.5cm]{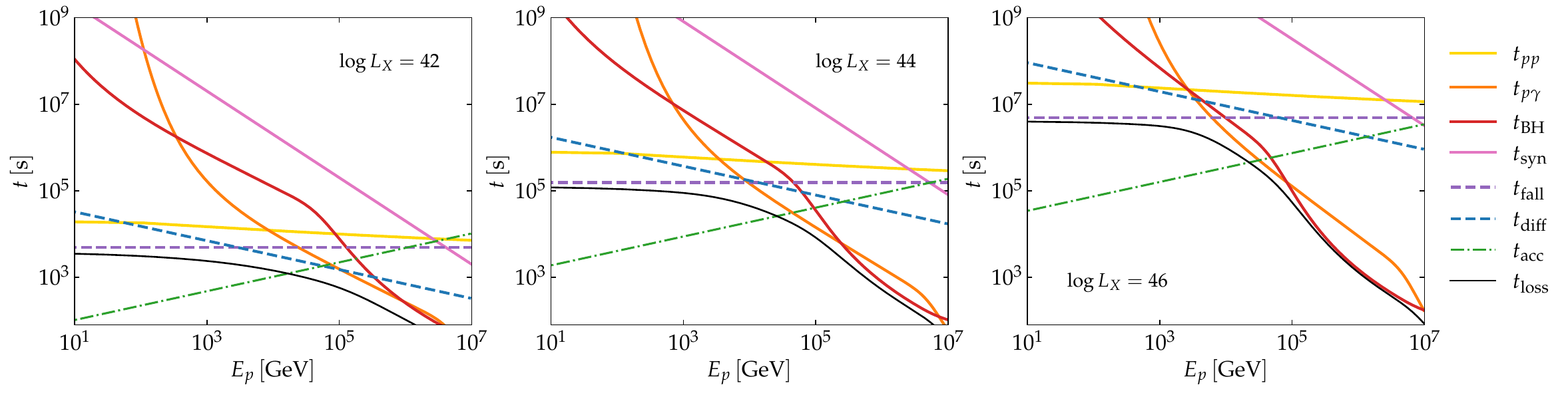}
\caption{Proton timescales for different X-ray luminosities of $L_X = 10^{42}\,\mathrm{erg}/\mathrm{s}$ (left panel), $10^{44}\,\mathrm{erg}/\mathrm{s}$ (middle panel), and $10^{46}\,\mathrm{erg}/\mathrm{s}$ (right panel). The inverse turbulence strength is set to $\eta = 56$ and the dimensionless coronal radius to $r_c = 5$. The thin black line corresponds to the total loss timescale defined in Eq.~(\ref{eq:t_loss}).}
\label{fig:proton_timescales}
\end{figure*}

Fig.~\ref{fig:proton_timescales} shows the proton interaction and escape timescales as well as the acceleration timescale for different X-ray luminosities, an inverse turbulence strength of $\eta = 56$, which corresponds to the best-fit value found for NGC 1068 in Section \ref{sec:fit}, and a coronal radius of $r_c = 5$. Sources with higher X-ray luminosities have a larger corona and, thus, a weaker magnetic field and lower densities of target protons and photons (see Table \ref{tab:corona_properties}). This leads to overall larger interaction and escape timescales and a slower acceleration process. As mentioned in Section \ref{sec:losses}, the cutoff energy of the proton spectrum is calculated as the energy at which the acceleration timescale is equal to the total loss timescale. This corresponds to the energy at which the green dash-dotted line and the thin black line in Fig.~\ref{fig:proton_timescales} intersect. For intermediate values of the inverse turbulence strength, like $\eta = 56$, the cutoff energy derived in this way does not exhibit a strong dependence on $L_X$. For sources with $L_X \gtrsim 10^{44}\,\mathrm{erg}/\mathrm{s}$, particle acceleration is mainly stopped by $p\gamma$ and Bethe-Heitler interactions, while for lower luminosities, escape via diffusion becomes equally important.

Due to the relatively small coronal radius of $r_c = 5$, the infall timescale is slightly shorter than the $pp$ interaction timescale at all energies and for all luminosities. Nevertheless, both $pp$ and $p\gamma$ interactions are important for the production of neutrinos, with $p\gamma$ interactions becoming more relevant for higher $L_X$. Energy losses via synchrotron radiation are negligible for all considered X-ray luminosities.

% -------------------------------------------------------------------------------------------------

\section{Sensitive energy range of the model fit for NGC 1068}
\label{sec:sensitive_energy_range}

In order to determine the sensitive energy range of the fit of our neutrino emission model for NGC 1068 to the public IceCube data, we estimate the central 68\% interval of neutrino energies that contribute the most to the excess of events over the background \cite{Schoenen2017, Bellenghi2024, Kontrimas2025}.

To this end, we select all detected neutrino events within a spatial box of size $\pm 10^\circ$ around the position of NGC 1068 and use \texttt{SkyLLH} \cite{Wolf2019, Bellenghi2023} to determine the contribution of each of these events to the test statistic (TS). The latter is defined as 
\begin{eqnarray}
\mathrm{TS} && \equiv -2 \log \frac{\mathcal{L}(n_s = 0)}{\mathcal{L}(\hat{n}_s, \hat{\eta})} \nonumber\\ && = 2 \sum_{i=1}^N \log \left[ \frac{\hat{n}_s}{N} \left( \frac{\mathcal{S}(\hat{E}_i, \hat{\theta}_i | \hat{\eta})}{\mathcal{B}(\hat{E}_i, \hat{\theta}_i)} - 1 \right) + 1 \right] ,
\end{eqnarray}
where $\mathcal{L}$ is the likelihood given by Eq.~(\ref{eq:likelihood}). To map the reconstructed muon energy, $\hat{E}$, to the true neutrino energy, $E_\nu$, we simulate neutrino events from a point source at the position of NGC 1068 and create a sample of $1.2 \times 10^8$ events with a flat distribution in $\log E_\nu$ between $100\,\mathrm{GeV}$ and $1\,\mathrm{EeV}$. This ensures sufficient statistics across the entire energy range. For each detected event, we select all simulated events whose reconstructed energies and angular uncertainties differ by less than 10\% from those of the detected event and construct the corresponding distribution in true neutrino energy \cite{Bellenghi2024}. Then, we compute the weighted sum of these distributions over all detected events, using their individual contributions to the test statistic as weights, and determine the central 68\% energy range from the resulting distribution. For our best-fit model neutrino spectrum of NGC 1068, this yields a sensitive energy range of $350\,\mathrm{GeV} - 190\,\mathrm{TeV}$.

\begin{figure}[t]
\includegraphics[width=8.5cm]{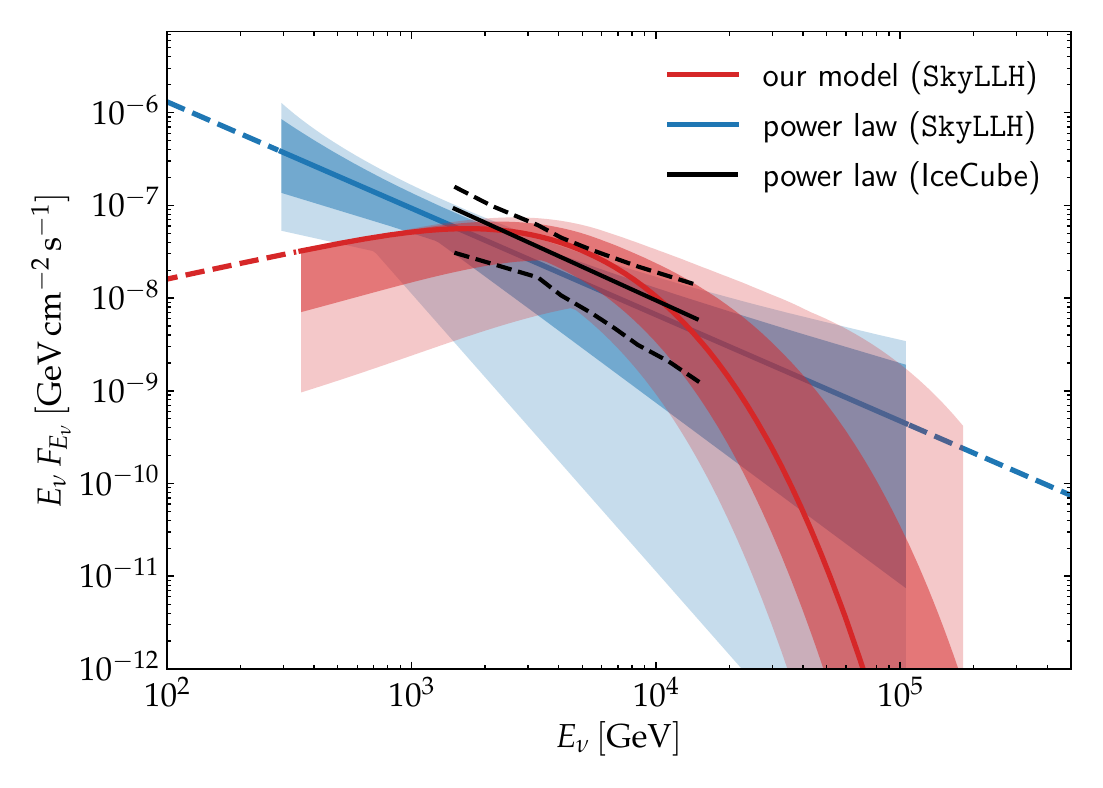}
\caption{Best-fit power-law neutrino spectrum of NGC 1068 (blue) together with our best-fit model neutrino spectrum (red). The shaded bands correspond to the $1\sigma$ and $2\sigma$ uncertainty regions, and their extent along the \textit{x}-axis indicates the central 68\% energy range of the respective analysis. The $2\sigma$ uncertainty band of the power-law spectrum was calculated with the spectral index restricted to $1.5\leq \gamma \leq 5.0$. The black solid line shows the best-fit power-law spectrum for NGC 1068 reported in \cite{Abbasi2022}, with the dashed lines indicating its 95\% confidence region. All fluxes are shown as all-flavor neutrino fluxes.}
\label{fig:comparison_our_model_pl} 
\end{figure}

In addition, we use \texttt{SkyLLH} to perform a fit of a simple power-law spectrum for NGC 1068 to the public IceCube data. In agreement with the results presented in \cite{Bellenghi2023}, we obtain $\hat{n}_s = 56.7$ signal events and a spectral index of $\hat{\gamma} = 3.2$, with the sensitive energy range of the analysis extending from $290\,\mathrm{GeV}$ to $110\,\mathrm{TeV}$. The resulting best-fit power-law spectrum is shown in Fig.~\ref{fig:comparison_our_model_pl} together with our best-fit model neutrino spectrum for NGC 1068. Overall, the results of the two analyses are compatible with each other. At energies $\gtrsim 500\,\mathrm{GeV}$, the spectra agree within their $2\sigma$ uncertainties, and at energies $\gtrsim 1\,\mathrm{TeV}$ within their $1\sigma$ uncertainties. However, the accuracy of both analyses is limited by the coarse binning of the publicly available instrument response function (IRF) in true neutrino energy and declination \cite{Abbasi2021}. This is the reason why the result of the power-law fit performed with \texttt{SkyLLH} differs from that reported in \cite{Aartsen2020b}, although the underlying event sample is the same. The limited energy resolution of the IRF also explains why we obtain relatively large sensitive energy ranges for both analyses.

% -------------------------------------------------------------------------------------------------

\section{Diffuse neutrino flux for different distributions of $P_\mathrm{CR}/P_\mathrm{th}$ and $\eta$}
\label{sec:diff_flux_distributions}

In the calculation of the diffuse neutrino flux in Section \ref{sec:results}, we assume that all sources share the same values of the pressure ratio and $\eta$. However, under this assumption, the neutrino emission from NGC 1068 cannot be explained without the diffuse neutrino flux produced by the entire population of Seyfert galaxies exceeding the observations by IceCube. Therefore, in a more realistic scenario, both parameters will likely follow some distribution among the population.

\begin{figure*}[b]
\includegraphics[width=17.5cm]{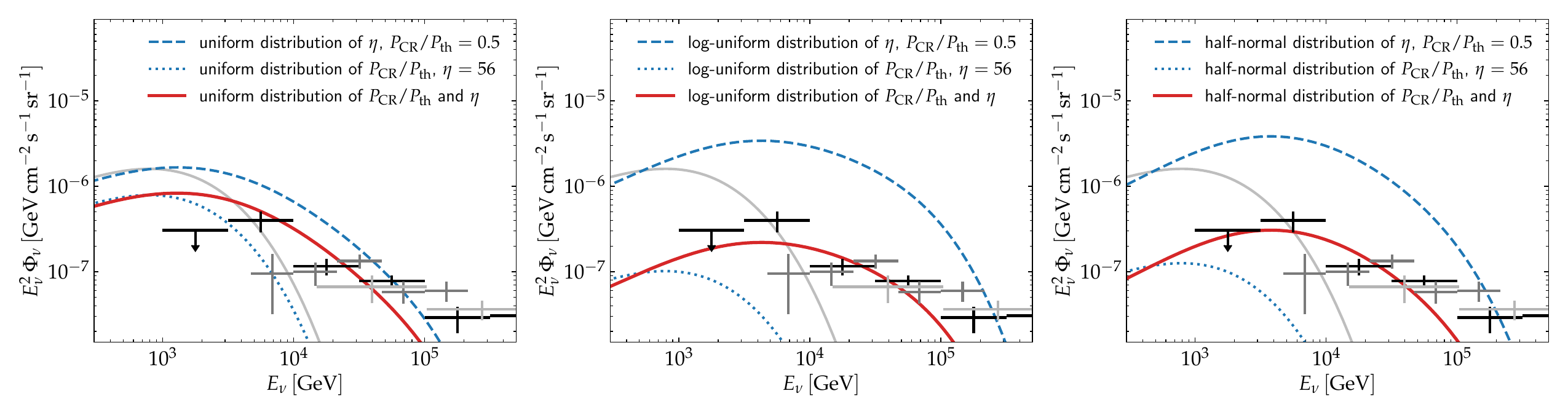}
\caption{Diffuse all-flavor neutrino flux for a uniform (left panel), a log-uniform (middle panel), and a half-normal distribution (right panel) of $P_\mathrm{CR}/P_\mathrm{th}$ and $\eta$ among the population, calculated for the U14 XLF. Red lines show the flux for the case in which both parameters follow the given distribution. Dashed (dotted) blue lines correspond to the case where only $\eta$ ($P_\mathrm{CR}/P_\mathrm{th}$) varies, while $P_\mathrm{CR}/P_\mathrm{th} = 0.5$ ($\eta = 56$) is fixed for all sources. For comparison, the flux obtained when fixing both parameters to the best-fit values for NGC 1068 is shown in gray. The black, dark gray, and light gray data points represent the diffuse neutrino flux measured by IceCube using starting tracks \cite{Abbasi2024c}, cascade events \cite{Aartsen2020}, and Northern sky tracks \cite{Abbasi2022b}, respectively. The parameters of the different distributions are given in Table \ref{tab:params_distributions}.}
\label{fig:diff_flux_distributions}
\end{figure*}

Fig.~\ref{fig:diff_flux_distributions} shows the diffuse neutrino flux predicted by our model for three possible distributions of $P_\mathrm{CR}/P_\mathrm{th}$ and $\eta$: a uniform distribution (left panel), a log-uniform distribution (middle panel), and a half-normal distribution (right panel). The parameters used for the different distributions are given in Table \ref{tab:params_distributions}. In general, changing from a fixed value of $\eta$ across the population to a distribution affects both the cutoff energy and the normalization of the diffuse neutrino flux (see blue dashed lines). The higher the abundance of sources with small values of $\eta$, the higher the cutoff energy and the normalization of the resulting spectrum. This is due to the fact that the neutrino spectra of sources with a higher level of turbulence not only extend to higher energies, but also have a higher overall normalization than those of sources with a less turbulent magnetic field (see Fig.~\ref{fig:neutrino_spectra}). In contrast, assuming a distribution of the pressure ratio instead of a fixed value for all sources in the population only affects the normalization, while the overall shape of the spectrum stays the same (see blue dotted lines).

The diffuse flux spectrum obtained for a uniform distribution of both $P_\mathrm{CR}/P_\mathrm{th}$ and $\eta$ has a slightly lower normalization and a higher cutoff energy compared to that for fixed parameter values. However, it still exceeds the IceCube observations at energies ${\lesssim}\,10\,\mathrm{TeV}$, suggesting that this distribution is unlikely to represent the actual underlying one. This is different for the other two distributions, both of which peak at low values of $P_\mathrm{CR}/P_\mathrm{th}$ and $\eta$ and lead to diffuse neutrino spectra that are broadly consistent with the IceCube flux measurements up to ${\sim}\,100\,\mathrm{TeV}$. Here, we have chosen the minimum value of $P_\mathrm{CR}/P_\mathrm{th}$ for the log-uniform distribution as well as the widths of the half-normal distributions of $P_\mathrm{CR}/P_\mathrm{th}$ and $\eta$ such that the resulting contribution of the Seyfert population to the observed diffuse flux is maximized without exceeding the upper limit at ${\sim}\,2\,\mathrm{TeV}$.

Overall, a distribution for which most sources have a low pressure ratio ($P_\mathrm{CR}/P_\mathrm{th} \lesssim 0.05$) and a high level of turbulence ($\eta \lesssim 10$) is necessary for the population of Seyfert galaxies to account for the observed diffuse neutrino flux up to energies of ${\sim}\,100\,\mathrm{TeV}$. In such a scenario, NGC 1068 with $P_\mathrm{CR}/P_\mathrm{th} = 0.5$ and $\eta = 56$ would stand out as an exceptional source, regardless of the exact form of the underlying distribution.

\onecolumngrid

\twocolumngrid

\begin{table*}
\caption{Parameters of the different distributions of $P_\mathrm{CR}/P_\mathrm{th}$ and $\eta$ underlying the diffuse flux spectra shown in Fig.~\ref{fig:diff_flux_distributions}. Listed are the lower and upper bounds of the considered parameter ranges as well as the mean and the standard deviation for the normal distribution.}
\begin{ruledtabular}
\begin{tabular}{lcccccccc}
Distribution & $(P_\mathrm{CR}/P_\mathrm{th})_\mathrm{min}$ & $(P_\mathrm{CR}/P_\mathrm{th})_\mathrm{max}$ & $\eta_\mathrm{min}$ & $\eta_\mathrm{max}$ & $\mu_{P_\mathrm{CR}/P_\mathrm{th}}$ & $\sigma_{P_\mathrm{CR}/P_\mathrm{th}}$ & $\mu_\eta$ & $\sigma_\eta$ \\
\hline
Uniform & 0.0 & 0.5 & 1 & 100 & -- & -- & -- & -- \\
Log-uniform & $10^{-7}$ & 0.5 & 1 & 100 & -- & -- & -- & -- \\
Half-normal & 0.0 & 0.5 & 1 & 100 & 0.0 & 0.05 & 1 & 10 \\
\end{tabular}
\end{ruledtabular}
\label{tab:params_distributions}
\end{table*}

% -------------------------------------------------------------------------------------------------

\bibliography{references.bib}

\end{document}